\documentclass[aps,prc,floatfix,twocolumn,superscriptaddress]{revtex4-1}

\usepackage[]{graphicx}
\usepackage{color}

\usepackage{amsmath,amssymb,amsfonts}
\usepackage{hyperref}
\hypersetup{
  colorlinks=true,        
  linkcolor=blue,         
  citecolor=cyan,         
}

\usepackage{amsmath,amssymb,amsfonts}

\usepackage{slashed}

\begin{document}

\title{Study of bottom quark dynamics via non-prompt $D^0$ and $J/\psi$ in Pb+Pb collisions at $\sqrt{s_\mathrm{NN}}=5.02$~TeV }

\author{Wen-Jing Xing}
\affiliation{Institute of Frontier and Interdisciplinary Science, Shandong University, Qingdao, Shandong 266237, China}

\author{Shu-Qing Li}
\email{lisq79@jnxy.edu.cn}
\affiliation{School of Physics and Electronic Engineering, Jining University, Qufu, Shandong, 273155, China}

\author{Shanshan Cao}
\email{shanshan.cao@sdu.edu.cn}
\affiliation{Institute of Frontier and Interdisciplinary Science, Shandong University, Qingdao, Shandong 266237, China}

\author{Guang-You Qin}
\email{guangyou.qin@ccnu.edu.cn}
\affiliation{Institute of Particle Physics and Key Laboratory of Quark and Lepton Physics (MOE), Central China Normal University, Wuhan, Hubei, 430079, China}

\date{\today}
\begin{abstract}

We study bottom quark energy loss via the nuclear modification factor ($R_\mathrm{AA}$) and elliptic flow ($v_2$) of non-prompt $D^0$ and $J/\psi$ in relativistic heavy-ion collisions at the LHC.
The space-time profile of quark-gluon plasma is obtained from the CLVisc hydrodynamics simulation, the dynamical evolution of heavy quarks inside the color deconfined QCD medium is simulated using a linear Boltzmann transport model that combines Yukawa and string potentials of heavy-quark-medium interactions, the hadronization of heavy quarks is performed using a hybrid coalescence-fragmentation model, and the decay of $B$ mesons is simulated via PYTHIA.
Using this numerical framework, we calculate the transverse momentum ($p_\mathrm{T}$) dependent $R_\mathrm{AA}$ and $v_2$ of direct $D$ mesons, $B$ mesons, and non-prompt $D^0$ and $J/\psi$ from $B$ meson decay in Pb+Pb collisions at $\sqrt{s_\mathrm{NN}}=5.02$~TeV.
We find the mass hierarchy of the nuclear modification of prompt $D$ and $B$ mesons depends on their $p_\mathrm{T}$. Both $R_\mathrm{AA}$ and $v_2$ of heavy flavor particles show strong $p_\mathrm{T}$ and centrality dependences due to the interplay between parton energy loss, medium geometry and flow, and hadronization of heavy quarks. Non-prompt $D^0$ and $J/\psi$ share similar patterns of $R_\mathrm{AA}$ and $v_2$ to $B$ mesons except for a $p_\mathrm{T}$ shift during the decay processes. Therefore, future more precise measurements on non-prompt $D^0$ and $J/\psi$ can help further pin down the bottom quark dynamics inside the quark-gluon plasma.

\end{abstract}
\maketitle

\section{Introduction}

The quark-gluon plasma (QGP), which consists of deconfined quarks and gluons as predicted by Quantum Chromodynamics (QCD) on the lattice, has been created in relativistic heavy-ion collisions performed at the Relativistic Heavy-Ion Collider (RHIC) and the Large Hadron Collider (LHC)~\cite{Gyulassy:2004zy, Muller:2012zq}.
Extensive studies have shown that the hot and dense QGP produced in these energetic nuclear collisions have two remarkable properties: small shear viscosity to entropy density ratio (or specific viscosity)~\cite{Romatschke:2017ejr, Rischke:1995ir, Heinz:2013th, Gale:2013da, Huovinen:2013wma},
and high opacity to the propagation of high-energy jet partons~\cite{Wang:1991xy, Gyulassy:2003mc, Majumder:2010qh, Qin:2015srf, Blaizot:2015lma, Cao:2020wlm, Cao:2022odi}.

Heavy quarks (charm and bottom quarks) are important hard probes of the QGP~\cite{Dong:2019byy, Andronic:2015wma, He:2022ywp}. They are mostly produced from the hard scatterings in the early stage of heavy-ion collisions, and then probe the entire history of the expanding QGP.
During their propagation through the QGP, heavy quarks interact with the medium and lose energy via collisional and radiative processes~\cite{Cao:2013ita, Cao:2016gvr, Xing:2019xae, Liu:2021dpm}.
The energy loss of heavy quarks in the QGP can be quantified by the nuclear modification factor $R_\mathrm{AA}$, defined as the ratio of the particle yield in a given centrality class in nucleus-nucleus (AA) collisions ($dN_\mathrm{AA}/dp_\mathrm{T}$), scaled by the average number of binary nucleon-nucleon collisions ($N_{\rm coll}$), to the particle yield in proton-proton ($pp$) collisions ($dN_{pp}/dp_\mathrm{T}$).
Experimental measurements have shown that the $R_\mathrm{AA}$ of heavy flavor particles produced in Au+Au collisions at RHIC and Pb+Pb collisions at the LHC is significantly below unity at high $p_\mathrm{T}$\cite{STAR:2014wif, CMS:2017qjw, STAR:2018zdy}, indicating substantial energy loss of heavy quarks inside the QGP medium due to their interactions with the medium constituents~\cite{Gossiaux:2006yu, Qin:2009gw, Uphoff:2011ad, Young:2011ug, Alberico:2011zy, Fochler:2013epa, Nahrgang:2013saa, Djordjevic:2013xoa, Cao:2015hia, Das:2015ana, Song:2015ykw, Kang:2016ofv, Prado:2016szr, Xu:2017obm, Liu:2017qah, Rapp:2018qla, Cao:2018ews, Li:2018izm, Ke:2018tsh, Li:2019wri, Katz:2019fkc, Li:2020kax, Chen:2021uar, Yang:2023rgb, Liu:2023rfi}.

Another important observable to probe the geometrical and dynamical properties of the QGP is the azimuthal anisotropy of the momentum space distribution of final state (soft and hard) hadrons~\cite{Ollitrault:1992bk, STAR:2000ekf, STAR:2001ksn, ALICE:2010suc, ALICE:2011ab}, which can be quantified by the Fourier coefficients of the particle distribution $dN/d\phi$ in the azimuthal plane.
These anisotropy coefficients can be calculated as $v_n =\langle \cos[n(\phi - \Phi_n)] \rangle$, with $\Phi_n$ being the $n$-th order event plane angle.
For example, $v_2$ is called elliptic flow, which mainly originates from the elliptic shape of the produced QGP matter.
Quantum fluctuations of the initial state (or colliding nuclei) can also contribute to the anisotropic flow, especially for odd flow harmonics and in ultra-central collisions~\cite{Alver:2010gr, Qin:2010pf}.
As for heavy quarks, due to their interactions with the anisotropic QGP, the final momentum distributions of heavy quarks and their daughter hadrons are also anisotropic.
At high $p_\mathrm{T}$, the heavy flavor hadron $v_2$ is sensitive to the energy loss difference of heavy quarks along different paths through the QGP.
At low and intermediate $p_\mathrm{T}$, their $v_2$ is sensitive to the collective flow of the QGP medium, since low $p_\mathrm{T}$ heavy quarks can pick up the QGP flow either by direct interactions with the medium through diffusive process or by coalescence with thermal light partons inside the medium during hadron formation~\cite{Moore:2004tg, He:2011qa, Xing:2021xwc, Plumari:2017ntm, He:2019vgs, Cho:2019lxb, Cao:2019iqs,Zhao:2023nrz}.
Experimental data have shown that $D$ mesons have positive $v_2$ at RHIC and the LHC~\cite{STAR:2017kkh, CMS:2017vhp, ALICE:2017pbx, ALICE:2020iug}, indicating charm quarks can build up significant collective flow through scatterings with an anisotropic QGP medium as well as the hadronization process.
The magnitude of the $D$ meson $v_2$ is similar to the light flavor hadron $v_2$ at intermediate $p_\mathrm{T}$, indicating the relaxation time of low $p_\mathrm{T}$ charm quarks might be comparable to or even shorter than the lifetime of the QGP.

In this work, we focus on bottom quark evolution in relativistic heavy-ion collisions.
Due to their even larger mass than charm quarks, bottom quarks provide a unique probe of the QGP properties.
First, they suffer much smaller cold nuclear matter effect than charm quarks~\cite{Kusina:2017gkz}, thus offer clean observables to study the hot nuclear matter effect.
In addition, bottom quarks provide a better tool to study the ``dead cone" effect which strongly depends on the mass-to-energy ratio ($m_Q/E$) of heavy quarks~\cite{Dokshitzer:2001zm, Armesto:2003jh, Zhang:2003wk, Djordjevic:2003zk, Zhang:2018nie}.
Therefore, at low to intermediate $p_\mathrm{T}$, bottom quarks are expected to experience less diffusion than charm quarks, and thus are harder to reach thermalization with QGP medium~\cite{Moore:2004tg,Xing:2021xwc,Liu:2023rfi}.
Recently, ALICE, ATLAS and CMS Collaborations have measured the $R_\mathrm{AA}$ and $v_2$ of non-prompt $D^0$ and $J/\psi$ decayed from bottom hadrons~\cite{CMS:2018bwt, ALICE:2022tji, ALICE:2023gjj, ATLAS:2018xms, ATLAS:2018hqe, ATLAS:2020yxw, ATLAS:2021xtw, ALICE:2019nuy, CMS:2023mtk, ALICE:2023hou}. Considering the large uncertainties of the current data directly on $B$ mesons, these measurements on non-prompt $D^0$ and $J/\psi$  provide an important supplementary opportunity for studying bottom quark interaction with the QGP in relativistic heavy-ion collisions. We will report our study on the $R_\mathrm{AA}$ and $v_2$ of bottom decayed $D^0$ and $J/\psi$, and compare them to results of prompt $D$ mesons directly produced from charm quark hadronization, in Pb+Pb collisions at $\sqrt{s_\mathrm{NN}}=5.02$~TeV from low to high $p_\mathrm{T}$ regimes.
The rest of this paper will be organized as follows. In Sec.~\ref{sec:model}, we will present our theoretical framework to study heavy quark evolution in relativistic heavy-ion collisions. The numerical results for prompt $D$ mesons, $B$ mesons, and non-prompt $D^0$ and $J/\psi$ will be presented and compared to available data at the LHC in Sec.~\ref{sec:results}. Section~\ref{sec:summary} will contain our summary.



\section{Theoretical framework: LBT-PNP model}
\label{sec:model}

In this work, we use our linear Boltzmann transport model that combines perturbative and non-perturbative interactions (LBT-PNP)~\cite{Xing:2021xwc} to simulate heavy quark scatterings through a color-deconfined medium.
In the LBT model~\cite{He:2015pra,Cao:2017hhk,Cao:2016gvr}, one solves the following Boltzmann equation for the evolution of the phase space distribution of heavy quarks (denoted by ``$a$") inside the QGP using the Monte-Carlo method:
\begin{eqnarray}
  P_a \cdot\partial f_a(x,p)=E_a (\mathcal{C}_\mathrm{el}+\mathcal{C}_\mathrm{inel}).
  \label{eq:boltzmann1}
\end{eqnarray}
The right hand side includes the contributions from both elastic and inelastic scatterings between heavy quarks and constituent partons of the medium, as denoted by the collision integrals $\mathcal{C}_\mathrm{el}$ and $\mathcal{C}_\mathrm{inel}$, respectively.

To simulate elastic scatterings between heavy quarks and medium partons, one calculates the scattering rates $\Gamma_{ab \rightarrow cd}$ for a binary collision process $a+b\to c+d$ using the following formula:
\begin{align}
  \label{eq:gamma_el}
   \Gamma_{ab \rightarrow cd} & (\vec{p}_a, T) = \frac{\gamma_b}{2E_a}\int \frac{d^3 p_b}{(2\pi)^3 2E_b} \frac{d^3 p_c}{(2\pi)^3 2E_c} \frac{d^3 p_d}{(2\pi)^3 2E_d}\nonumber
  \\ \times\,& f_b (\vec{p}_b, T) [1 \pm f_c (\vec{p}_c, T)] [1 \pm f_d (\vec{p}_d, T)]\nonumber
  \\ \times\, &\theta (s - (m_a + \mu_d)^2) \nonumber
  \\ \times\,& (2\pi)^4 \delta^{(4)}(p_a + p_b - p_c -p_d) |\mathcal{M}_{ab \rightarrow cd}|^2.
\end{align}
In the above equation, $\gamma_b$ is the degeneracy factor of parton $b$, $f_b$ and $f_d$ are thermal distributions of the medium partons, $1\pm f$ is the Bose enhancement or Fermi suppression factor for the final states (neglected for heavy quark $c$ due to their dilute distribution in this work), and the $\theta$-function accounts for the thermal mass effect on medium partons, where $\mu_d$ represents the Debye mass.
In this calculation, the masses of charm and bottom quarks are taken as $M_c = 1.27$~GeV and $M_b = 4.19$~GeV, and the medium partons are taken to be massless.
The key information about the microscopic elastic scattering process is contained in the matrix element $\mathcal{M}_{ab \rightarrow cd}$.

In the LBT-PNP model~\cite{Xing:2021xwc}, the matrix element $\mathcal{M}_{ab \rightarrow cd}$ for elastic scatterings includes both perturbative and non-perturbative interactions between heavy quarks and medium constituents. More specifically, we use the following parameterized Cornell-type potential for the interaction between a heavy quark and a medium parton,
\begin{eqnarray}
\label{eq:Cornell_poten1}
V(r) = V_\mathrm{Y}(r) + V_\mathrm{S}(r) = -\frac{4}{3} \alpha_\mathrm{s} \frac{e^{-m_d r}}{r} - \frac{\sigma e^{-m_s r}}{m_s}.
\end{eqnarray}
One can see that the above potential includes both short-range Yukawa interaction and long-range color confining interaction, the later of which is also called the string term.
In these two terms, $\alpha_\mathrm{s}$ and $\sigma$ are the coupling strengths for the Yukawa and string interactions respectively, $m_d$ and $m_s$ are their corresponding screening masses, which are taken to be temperature dependent as $m_d = a + bT$ and $m_s = \sqrt{a_s + b_sT}$.
In this work, the values of the parameters $\alpha_\mathrm{s},\sigma, a,b, a_s, b_s$ are taken from Ref.~\cite{Xing:2021xwc}, which provides a reasonable description of the $D$ meson observables measured at RHIC and the LHC.

To calculate the matrix element $\mathcal{M}_{ab \rightarrow cd}$, one takes the Fourier transformation and obtains the above Cornell-type potential in the momentum space as
\begin{eqnarray}
\label{eq:HQ_potential}
V(\vec{q}) = - \frac{4\pi \alpha_\mathrm{s} C_F}{m_d^2+|\vec{q}|^2} - \frac{8\pi \sigma}{(m_s^2+|\vec{q}|^2)^2},
\end{eqnarray}
with $\vec{q}$ being the momentum exchange between heavy quarks and medium constituents.
To calculate the matrix element for two-body processes: $Qq\to Qq$ and $Qg\to Qg$, in which gluons with momentum $\vec{q}$ are exchanged, we treat the above in-medium Cornell-type potential as the effective gluon propagator (field).
Assuming a scalar interaction vertex for the string term, the scattering amplitude can be written as:
\begin{align}
  \label{eq:Matrix_cq}
 i \mathcal{M}  =\,& \mathcal{M_\mathrm{Y}} + \mathcal{M_\mathrm{S}}\nonumber
 \\=\,& \overline{u}(p') \gamma^{\mu} u(p) V_\mathrm{Y}(\vec{q}) \overline{u}(k') \gamma^{\nu} u(k)\nonumber
 \\&+ \overline{u}(p') u(p) V_\mathrm{S}(\vec{q}) \overline{u}(k') u(k).
\end{align}
Here, $\mathcal{M_\mathrm{Y}}$ and $\mathcal{M_\mathrm{S}}$ represent the matrix elements for the Yukawa and string terms respectively.
Note that we still use a vector interaction vertex for the Yukawa term, which can reproduce the leading-order perturbative QCD result~\citep{Combridge:1978kx}.
Since the in-medium potential represents the effective gluon propagator, the string term is only included for the $t$-channel scattering, i.e., by setting $|\vec{q}|^2 = -t$ in the above potential.
Note that the color information of interaction vertices has been included in the interaction potential. For the $Qq \rightarrow Qq$ scattering process, the final amplitude squared is given by:
\begin{align}
  \label{eq:M2_cq}
|\mathcal{M}_{Qq}|^2 &= \frac{64\pi^2 \alpha_\mathrm{s}^2}{9} \frac{(s-m_Q^2)^2+(m_Q^2-u)^2+2m_Q^2 t}{(t-m_d^2)^2}\nonumber\\
 &+ \frac{(8\pi \sigma)^2}{N_c^2 -1} \frac{t^2-4m_Q^2 t}{(t-m_s^2)^4};
\end{align}
and for the $Qg \rightarrow Qg$ process, we have
\begin{align}
  \label{eq:M2_cg}
  |\mathcal{M}&_{Qg}|^2 = \nonumber\\
  & \frac{64\pi^2 \alpha_\mathrm{s}^2}{9} \frac{(s-m_Q^2)(m_Q^2-u)+2m_Q^2 (s+m_Q^2)}{(s-m_Q^2)^2}\nonumber
  \\+\, & \frac{64\pi^2 \alpha_\mathrm{s}^2}{9} \frac{(s-m_Q^2)(m_Q^2-u)+2m_Q^2 (u+m_Q^2)}{(u-m_Q^2)^2}\nonumber
  \\+\, & 8\pi^2 \alpha_\mathrm{s}^2 \frac{5m_Q^4 + 3m_Q^2t -10m_Q^2u + 4t^2 + 5tu + 5u^2}{(t-m_d^2)^2}\nonumber
  \\+\, & 8\pi^2 \alpha_\mathrm{s}^2 \frac{(m_Q^2-s)(m_Q^2-u)}{(t-m_d^2)^2}\nonumber
  \\+\, & 16\pi^2 \alpha_\mathrm{s}^2 \frac{3m_Q^4 -3m_Q^2s - m_Q^2u + s^2}{(s-m_Q^2)(t-m_d^2)}\nonumber
  \\+\, & \frac{16\pi^2 \alpha_\mathrm{s}^2}{9} \frac{m_Q^2 (4m_Q^2 - t)}{(s - m_Q^2)(m_Q^2 - u)}\nonumber
  \\+\, & 16\pi^2 \alpha_\mathrm{s}^2 \frac{3m_Q^4 - m_Q^2s -3m_Q^2u +u^2}{(t - m_d^2)(u - m_Q^2)}\nonumber
  \\+\, & \frac{C_A}{C_F} \frac{(8\pi \sigma)^2}{N_c^2 -1} \frac{t^2-4m_Q^2 t}{(t-m_s^2)^4},
\end{align}
with $s$, $t$, $u$ being the Mandelstam variables. To obtain the Yukawa parts $|\mathcal{M_\mathrm{Y}}|^2$ in the above two expressions, we calculate $|\mathcal{M}_s|^2$, $|\mathcal{M}_u|^2$, $|\mathcal{M}_t|^2$, and their interference terms $|\mathcal{M}_{s} \mathcal{M}_t^{*}|$, $|\mathcal{M}_{s} \mathcal{M}_u^{*}|$ and $|\mathcal{M}_{u} \mathcal{M}_t^{*}|$ separately, and then replace $t$ by $t-m_d^2$ in the denominators. Note that the form of the $|\mathcal{M_\mathrm{Y}}|^2$ part in Eq.~(\ref{eq:M2_cg}) appears a little different from the form given by Ref.~\citep{Combridge:1978kx}. This is because the amplitude squared for the $Qg \rightarrow Qg$ scattering process is simplified in Ref.~\citep{Combridge:1978kx} by extracting a term $-16\pi^2 \alpha_\mathrm{s}^2 (2m_Q^2+t)/t$ from both $|\mathcal{M}_{s} \mathcal{M}_t^{*}|$ and $|\mathcal{M}_{u} \mathcal{M}_t^{*}|$, and then combine them with $|\mathcal{M}_t|^2$ to obtain $32\pi^2 \alpha_\mathrm{s}^2 (s-m_Q^2)(m_Q^2-u)/t^2$. If one introduces the Debye screening mass, i.e, replace $t$ by $t-m_d^2$ in the denominators, based on the form in Ref.~\citep{Combridge:1978kx}, as shown in Refs.~\citep{Svetitsky:1987gq,GolamMustafa:1997id,Liu:2016ysz}, the matrix element would appear different from our Yukawa part in Eq.~(\ref{eq:M2_cg}). However, they are the same when $m_d$ is set as zero.

By summing over all possible scattering channels, one can obtain the total elastic scattering rate for a heavy quark propagating through the QGP according to Eq.~(\ref{eq:gamma_el}).
For a given time step $\Delta \tilde{t}$, the probability for a heavy quark to experience elastic scatterings with the QGP constituents can be calculated as: $P_\mathrm{el}^a = 1-e^{- \Gamma_\mathrm{el}^a \Delta \tilde{t}}$.

For inelastic scatterings between heavy quarks and the QGP, the scattering rate at a given time $\tilde{t}$ can be obtained as follows:
\begin{eqnarray}
  \label{eq:gamma_inel}
  \Gamma_\mathrm{inel}^a (E_a, T, \tilde{t}) = \int dxdl_{\bot}^2 \frac{dN_g^a}{dxdl_\bot ^2 d\tilde{t}}.
\end{eqnarray}
In this study, we take the higher-twist energy loss formalism~\cite{Guo:2000nz,Majumder:2009ge,Zhang:2003wk,Zhang:2018nie} for the gluon emission spectrum off a heavy quark inside a dense nuclear medium,
\begin{eqnarray}
  \label{eq:gluon_spectrum}
  \frac{dN_g^a}{dxdl_\bot ^2 d\tilde{t}} = \frac{2C_A \alpha_\mathrm{s} P_a(x) l_{\bot}^4 \hat{q}_a}{\pi (l_{\bot}^2 + x^2 m_a^2)^4} \sin^2 \left(\frac{\tilde{t}-\tilde{t}_i}{2\tau_f}\right).
\end{eqnarray}
In the above two equations, $E_a$ and $m_a$ are the energy and mass of heavy quarks, $x$ and $l_{\perp}$ represent the energy fraction and the transverse momentum of the radiated gluon with respect to the parent heavy quark, and $P_a(x)$ is the vacuum splitting function.
The jet transport coefficient $\hat{q}_a$ represents the average transverse momentum squared exchanged between heavy quarks and the medium constituents per unit time during the elastic scattering process.
Inside the sine function, $\tilde{t}-\tilde{t}_i$ is the time accumulated from the previous emission time ($\tilde{t}_i$), and $\tau_f = 2E_a x(1-x)/(l_{\bot}^2+x^2m_a^2)$ denotes the formation time of gluon emission.
Based on the above formula, the probability for a heavy quark to experience inelastic scatterings with the QGP constituents during a time step $\Delta \tilde{t}$ can be calculated as: $P_\mathrm{inel}^a = 1-e^{-\Gamma_\mathrm{inel}^a \Delta \tilde{t}}$.

To include both elastic and inelastic scatterings between heavy quarks and the QGP medium, the total scattering rate can be obtained as: $\Gamma_\mathrm{tot}=\Gamma_\mathrm{el}+\Gamma_\mathrm{inel}$. In terms of the total scattering probability, one may write out the following expression:
\begin{eqnarray}
  \label{eq:prob_total}
  P_\mathrm{tot}^a = 1 - e^{-\Gamma_\mathrm{tot} \Delta t} = P_\mathrm{el}^a + P_\mathrm{inel}^a - P_\mathrm{el}^a P_\mathrm{inel}^a,
\end{eqnarray}
where  $P_\mathrm{el}^a (1-P_\mathrm{inel}^a)$ can be understood as the pure elastic scattering probability, and $P_\mathrm{inel}^a$ is the inelastic scattering probability.
Following our earlier work~\cite{Cao:2017hhk, Xing:2019xae, Xing:2021xwc}, we use different values of the coupling strength $\alpha_\mathrm{s}$ for different interaction vertices in our calculation of elastic and inelastic scatterings.
For a vertex connecting to the propagating heavy quarks, we take $\alpha_\mathrm{s} = 4\pi / [9 \mathrm{ln}(2ET/\Lambda^2)]$ with $\Lambda=0.2$~GeV; for a vertex connecting to the medium partons, we use the same value as in the interaction potential $V(r)$.

\begin{figure*}[tb]
\includegraphics[width=0.495\linewidth]{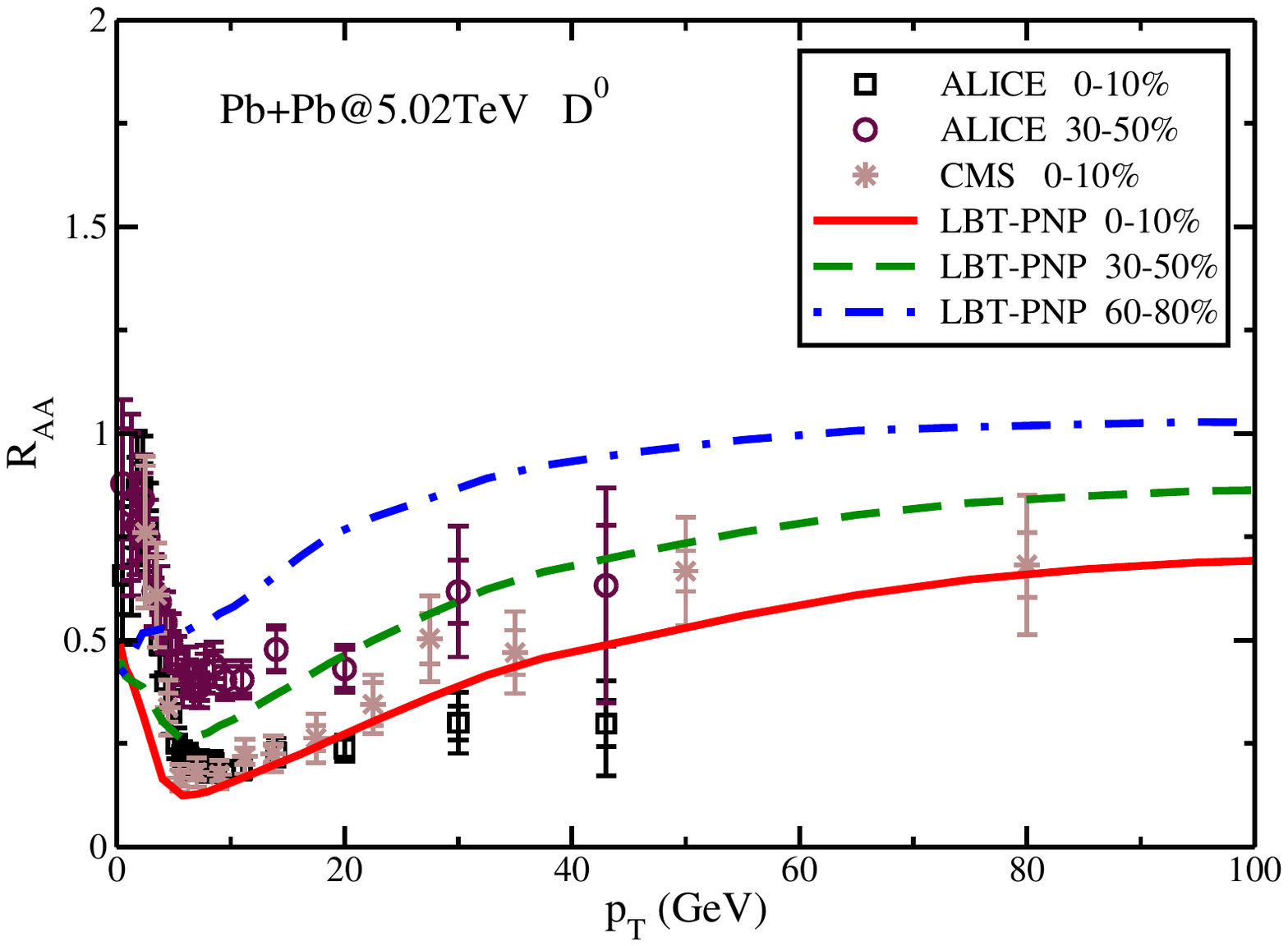}
\includegraphics[width=0.495\linewidth]{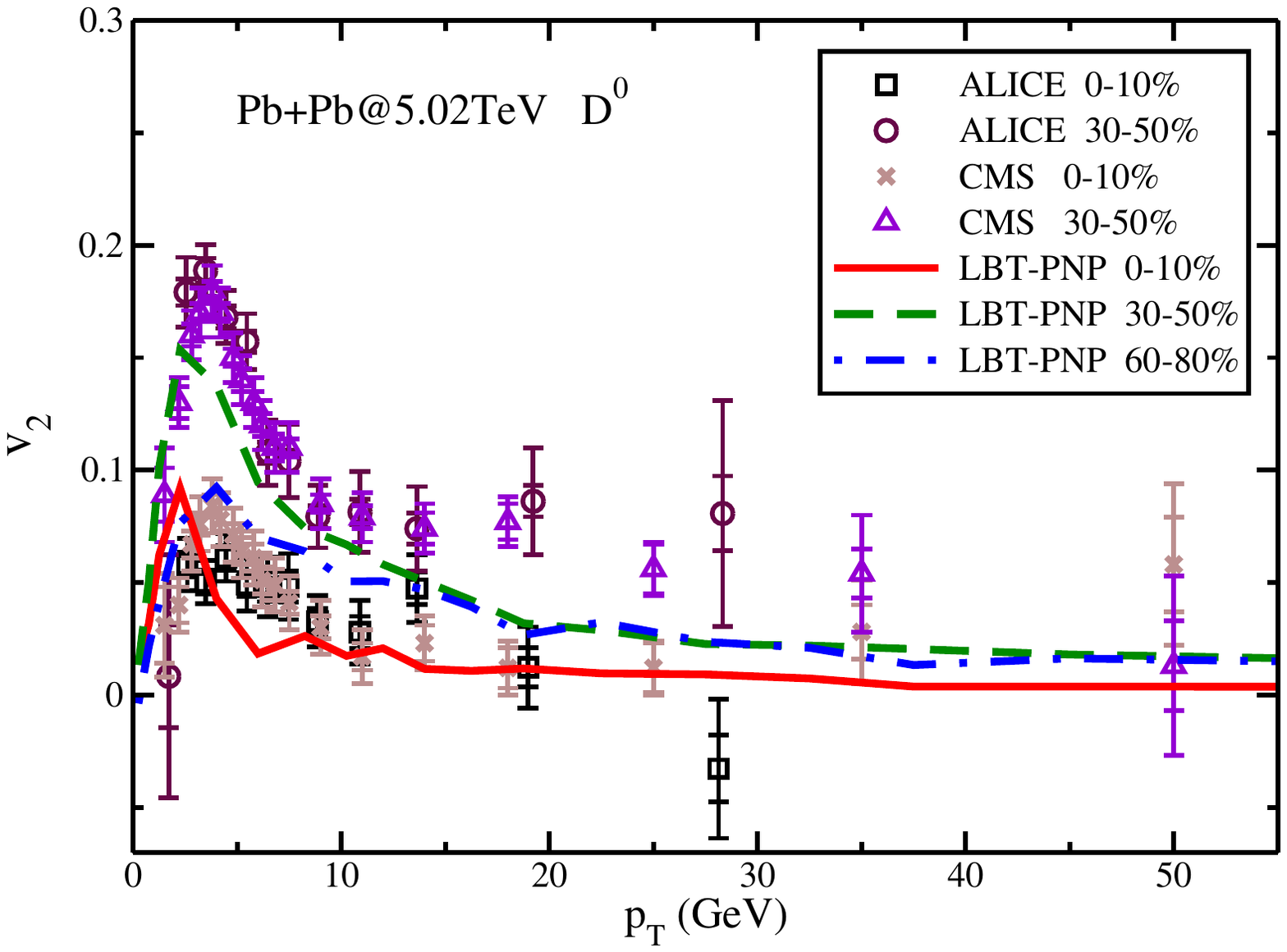}
\caption{(Color online) $R_\mathrm{AA}$ and $v_2$ of the prompt $D^0$ mesons as functions of $p_\mathrm{T}$ in 0-10\%, 30-50\% and 60-80\% Pb+Pb collisions at $\sqrt{s_\mathrm{NN}}=5.02$~TeV, compared to the ALICE and CMS data for 0-10\% and 30-50\% centralities~\cite{ALICE:2021rxa,CMS:2017qjw,ALICE:2020iug,CMS:2020bnz}.
} \label{fig1}
\end{figure*}

To evolve heavy quarks through a realistic medium, we utilize the (3+1)-dimensional viscous hydrodynamic model CLVisc~\cite{Pang:2012he,Pang:2018zzo,Wu:2018cpc,Wu:2021fjf} to simulate the dynamical evolution of the QGP fireball produced in relativistic heavy-ion collisions at the LHC. In the present study, we apply smooth hydrodynamic profiles for investigating heavy quarks, whose specific shear viscosity is taken to be $\eta/s=0.16$ for the QGP produced in Pb+Pb collision at $\sqrt{s_\mathrm{NN}} = 5.02$~TeV.
The initial energy density distribution of the QGP is obtained from the Glauber model, which is also used to obtain the initial spatial distribution of heavy quarks.
The momentum distribution of heavy quarks is initialized using the fixed-order-next-to-leading-log (FONLL) calculation~\cite{Cacciari:2001td,Cacciari:2012ny,Cacciari:2015fta}, where the parton distribution functions are taken from CT14NLO~\cite{Dulat:2015mca}, and the nuclear shadowing effect is taken from EPPS16~\cite{Eskola:2016oht} at the next-to-leading-order.
After their production, heavy quarks are assumed to stream freely before starting interaction with the QGP at the initial proper time of hydrodynamic evolution ($\tau_0=0.6$~fm/$c$).
Interaction between heavy quarks and the QGP ceases when heavy quarks reach the QGP boundary, the hypersurface with $T_{\rm c} = 165$~MeV in this work, on which they are converted into heavy flavor hadrons via a hybrid fragmentation-coalescence model~\cite{Cao:2019iqs}. The decay of $B$ mesons into non-prompt $D^0$ and $J/\psi$ is simulated by Pythia~\cite{Sjostrand:2006za}.

\begin{figure*}[tb]
\includegraphics[width=0.495\linewidth]{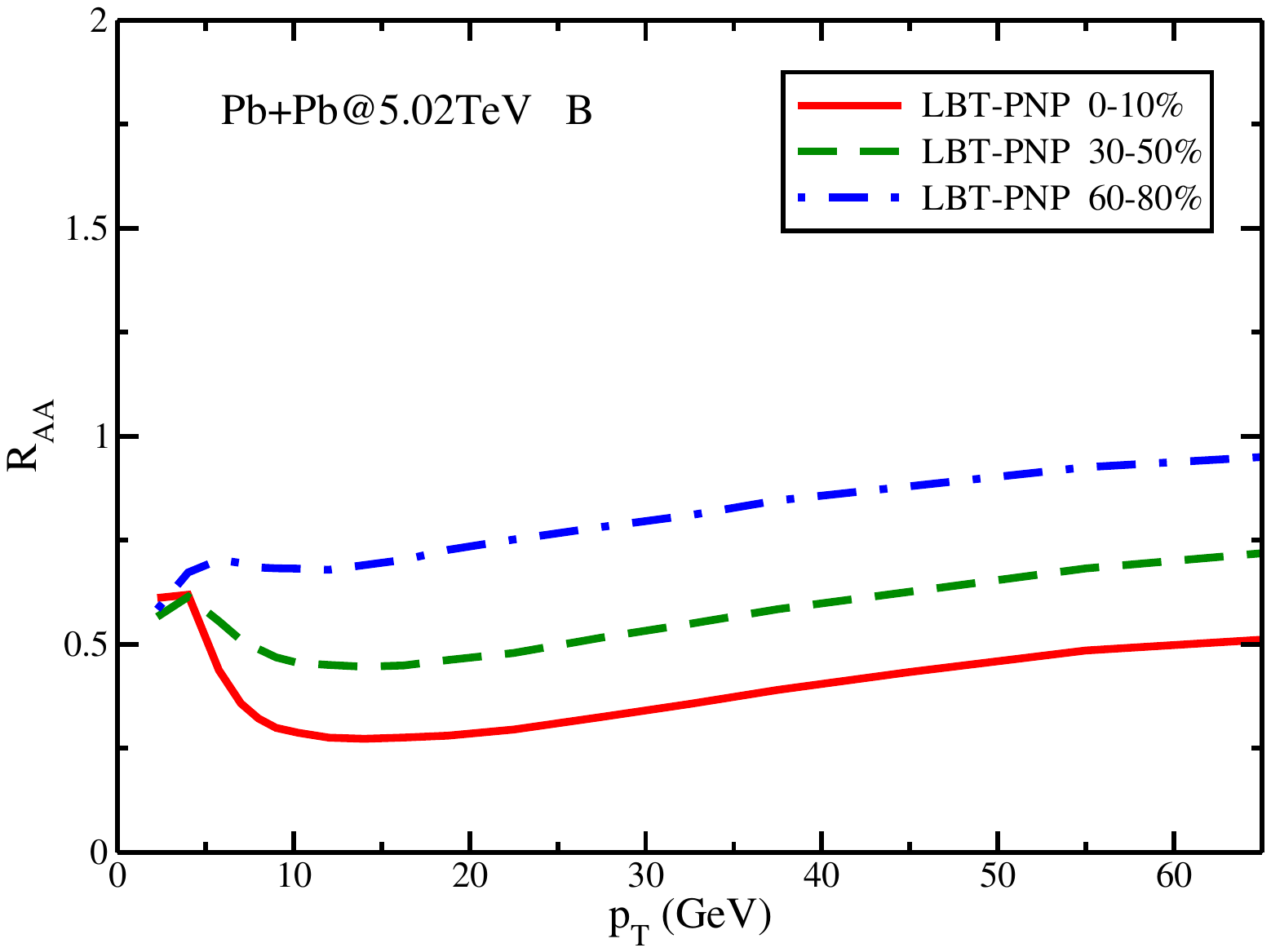}
\includegraphics[width=0.495\linewidth]{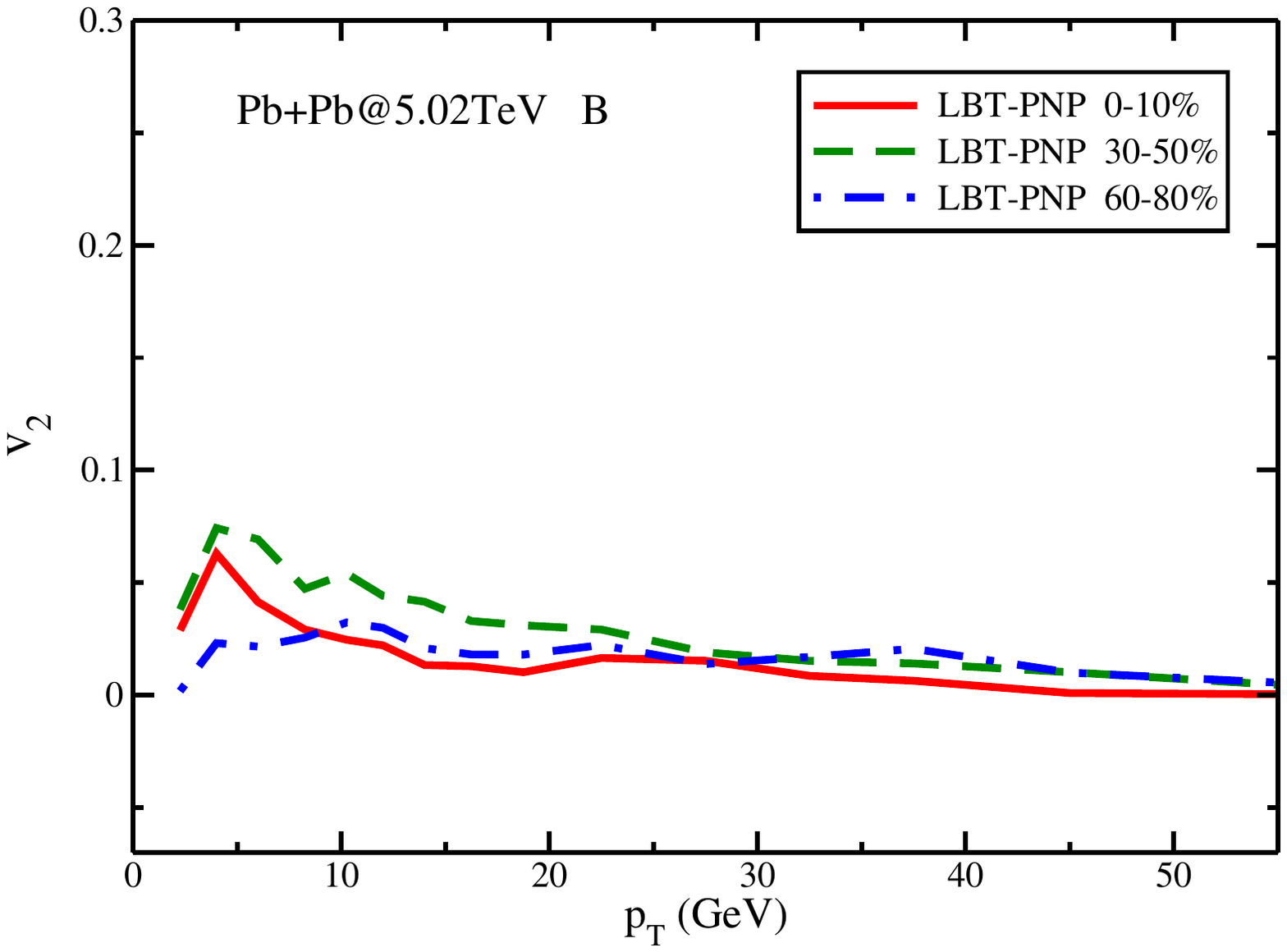}
\caption{(Color online) $R_\mathrm{AA}$ and $v_2$ of $B$ mesons as functions of $p_\mathrm{T}$ in 0-10\%, 30-50\% and 60-80\% Pb+Pb collisions at $\sqrt{s_\mathrm{NN}}=5.02$~TeV.
} \label{fig2}
\end{figure*}

\begin{figure*}[tb]
\includegraphics[width=0.495\linewidth]{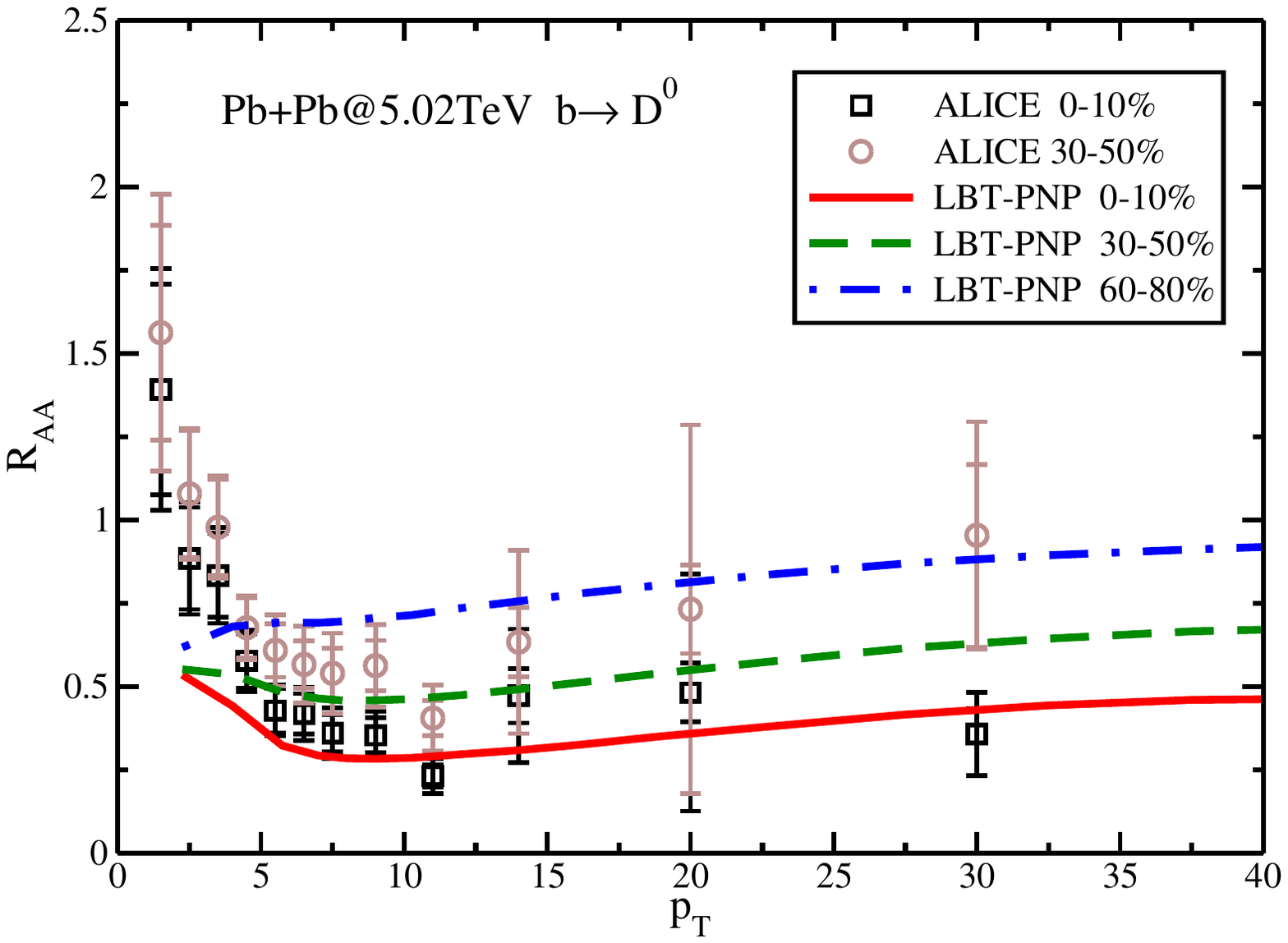}
\includegraphics[width=0.495\linewidth]{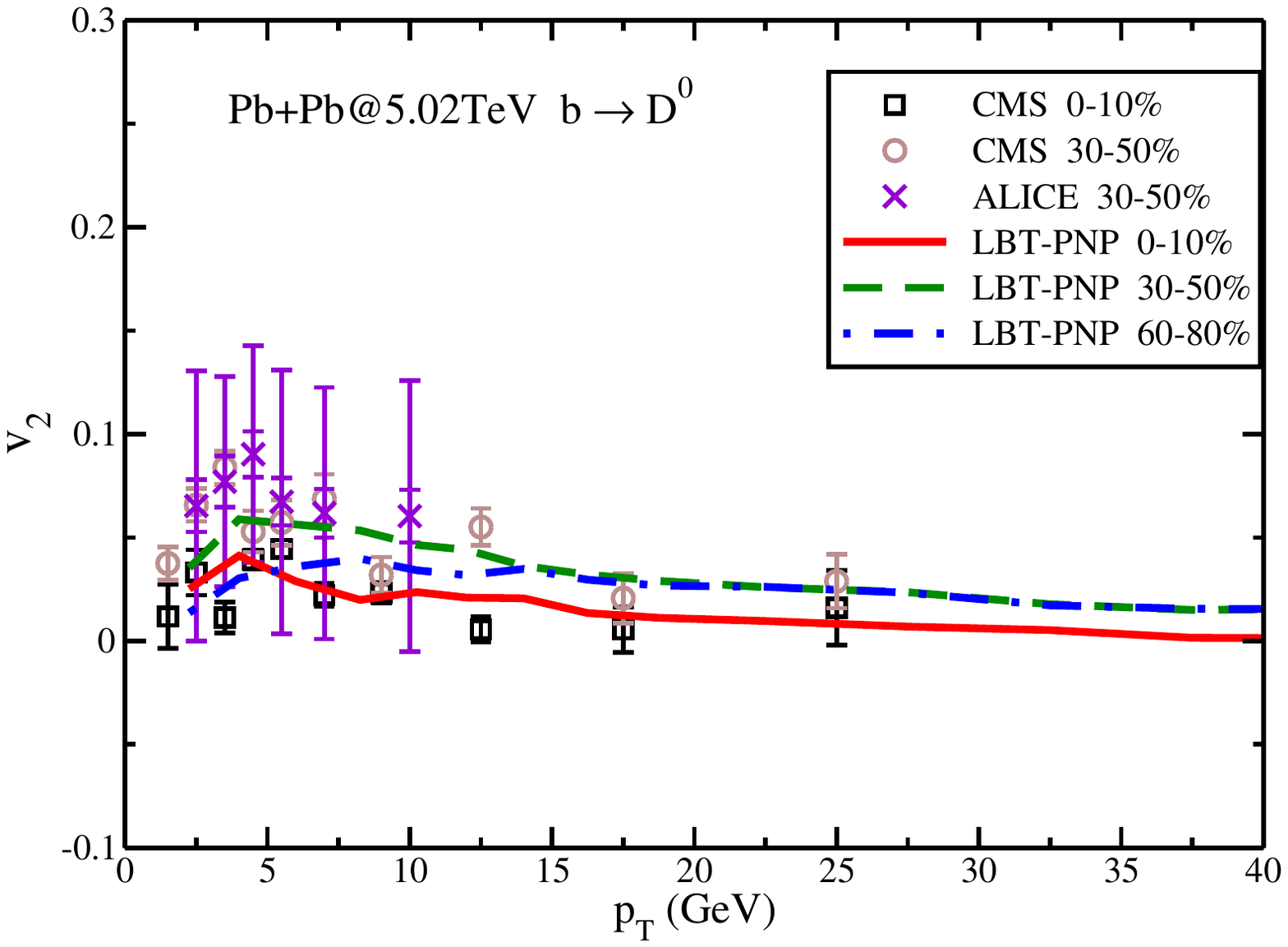}
\caption{(Color online) $R_\mathrm{AA}$ and $v_2$ of non-prompt $D^0$ as functions of $p_\mathrm{T}$ in 0-10\%, 30-50\% and 60-80\% Pb+Pb collisions at $\sqrt{s_\mathrm{NN}}=5.02$~TeV, compared to the ALICE data for 0-10\% and 30-50\% centralities~\cite{ALICE:2022tji,ALICE:2023gjj,CMS:2022vfn}.
} \label{fig3}
\end{figure*}

\begin{figure*}[tb]
\includegraphics[width=0.495\linewidth]{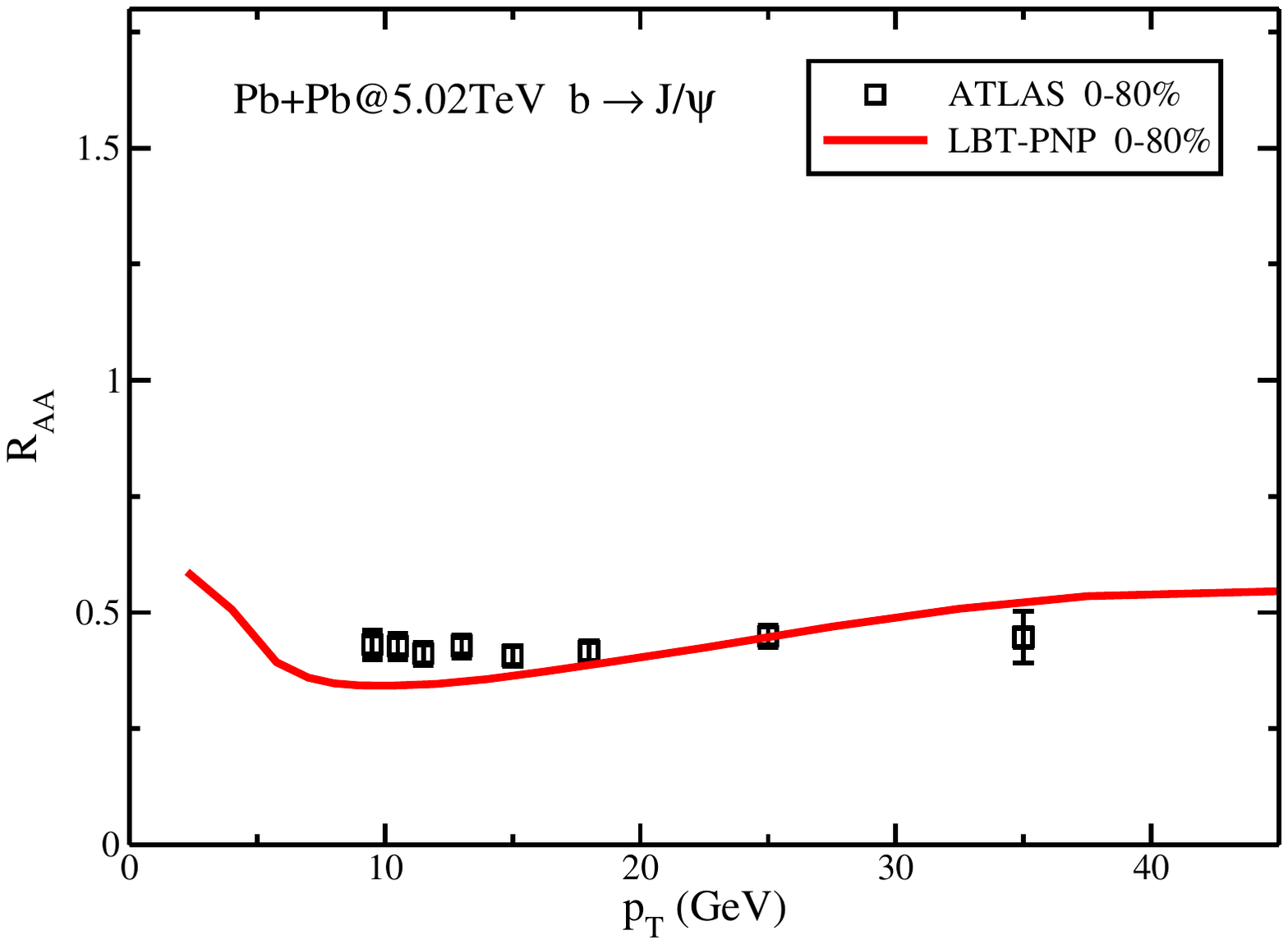}
\includegraphics[width=0.495\linewidth]{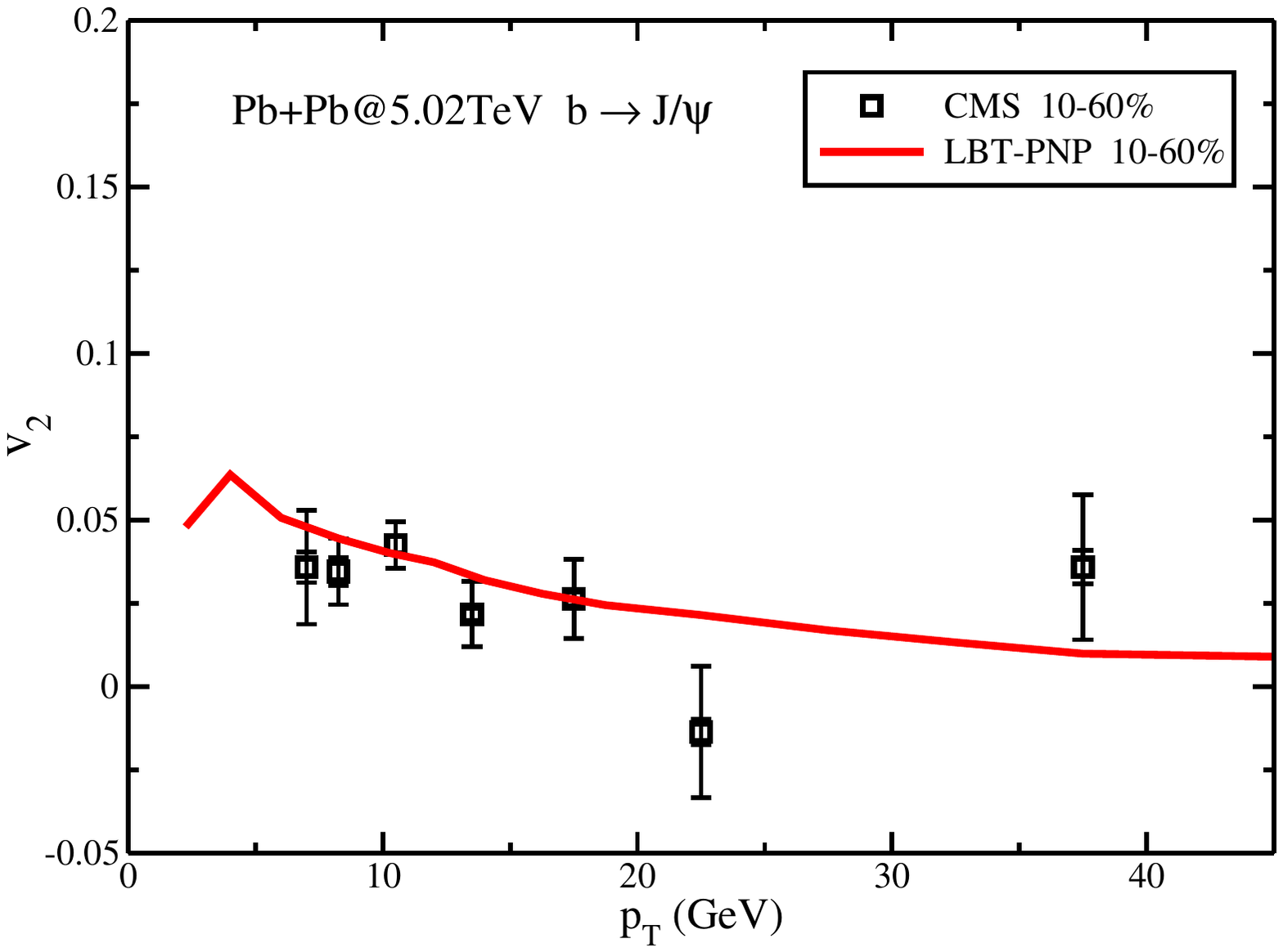}
\caption{(Color online) $R_\mathrm{AA}$ (0-80\% centrality) and $v_2$ (10-60\% centrality) of non-prompt $J/\Psi$ as functions of $p_\mathrm{T}$ in Pb+Pb collisions at $\sqrt{s_\mathrm{NN}}=5.02$~TeV, compared to the ATLAS and CMS data~\cite{ATLAS:2018hqe,CMS:2023mtk}.
} \label{fig4}
\end{figure*}

\begin{figure*}[tb]
\includegraphics[width=0.495\linewidth]{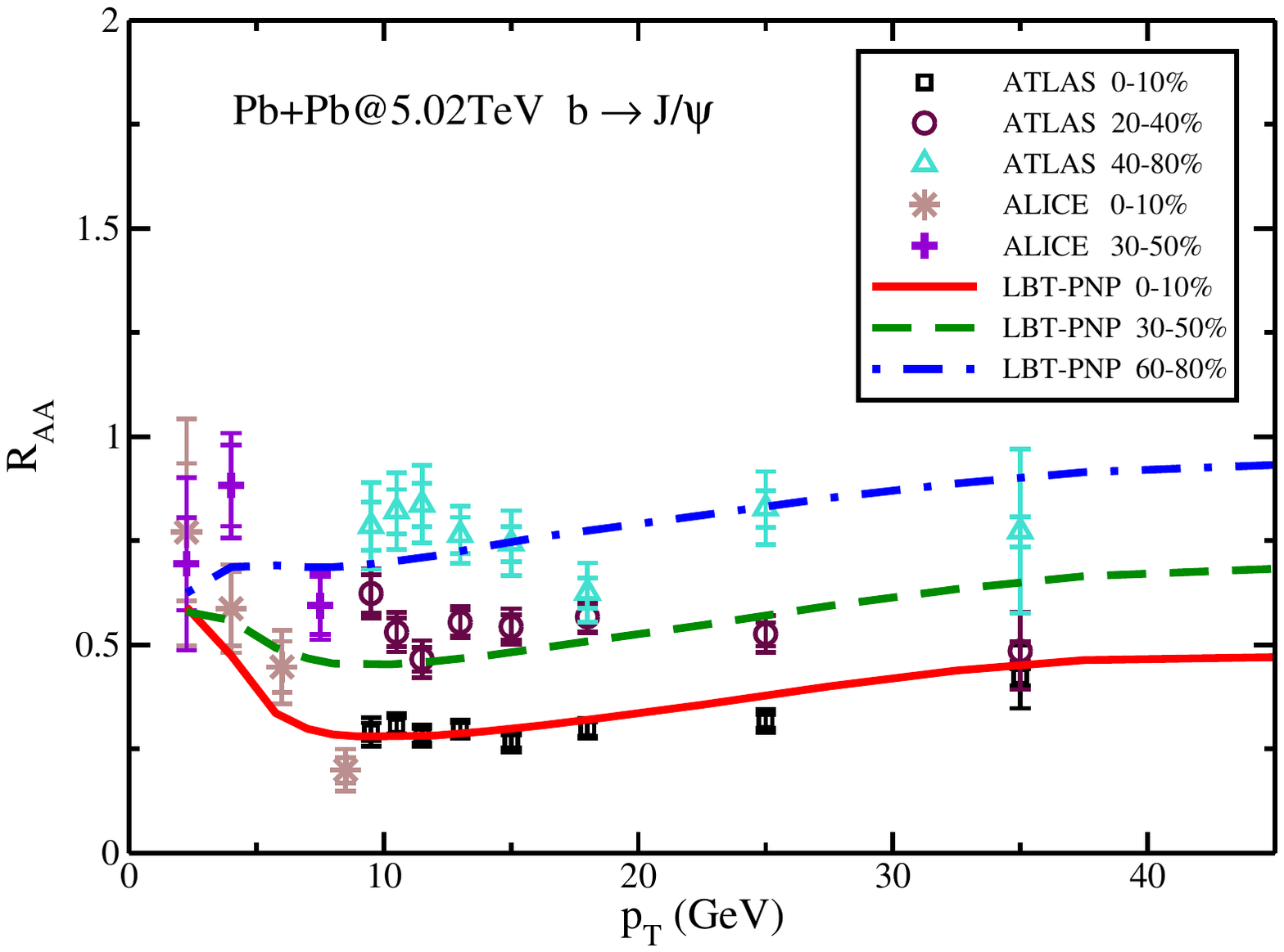}
\includegraphics[width=0.495\linewidth]{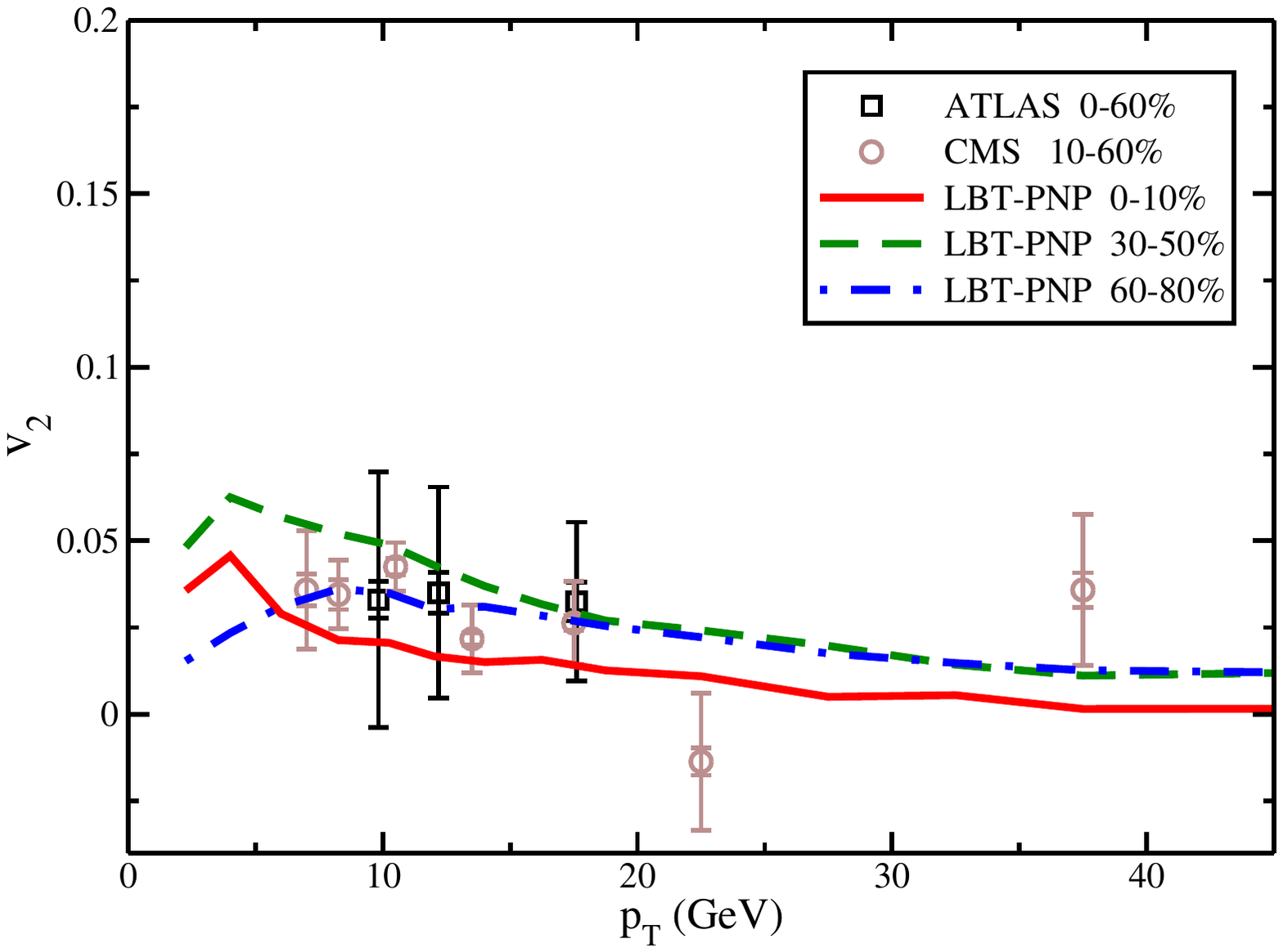}
\caption{(Color online) $R_\mathrm{AA}$ and $v_2$ of non-prompt $J/\Psi$ as functions of $p_\mathrm{T}$ in 0-10\%, 30-50\% and 60-80\% Pb+Pb collisions at $\sqrt{s_\mathrm{NN}}=5.02$~TeV, compared to the ALICE, ATLAS and CMS data for various centralities~\cite{ATLAS:2018hqe,ALICE:2023hou,CMS:2023mtk,ATLAS:2018xms}.
} \label{fig5}
\end{figure*}

\begin{figure*}[tb]
\includegraphics[width=0.495\linewidth]{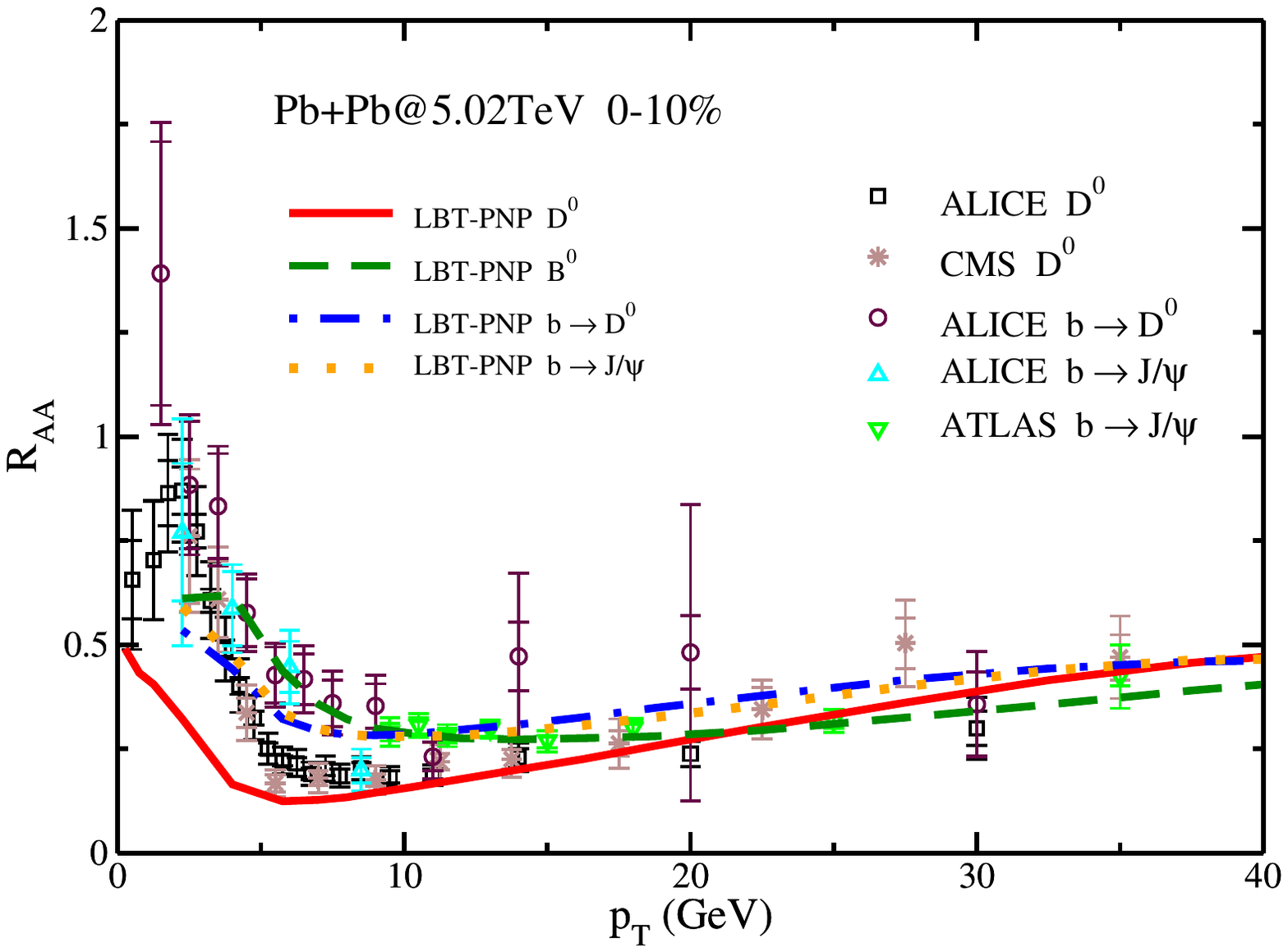}
\includegraphics[width=0.495\linewidth]{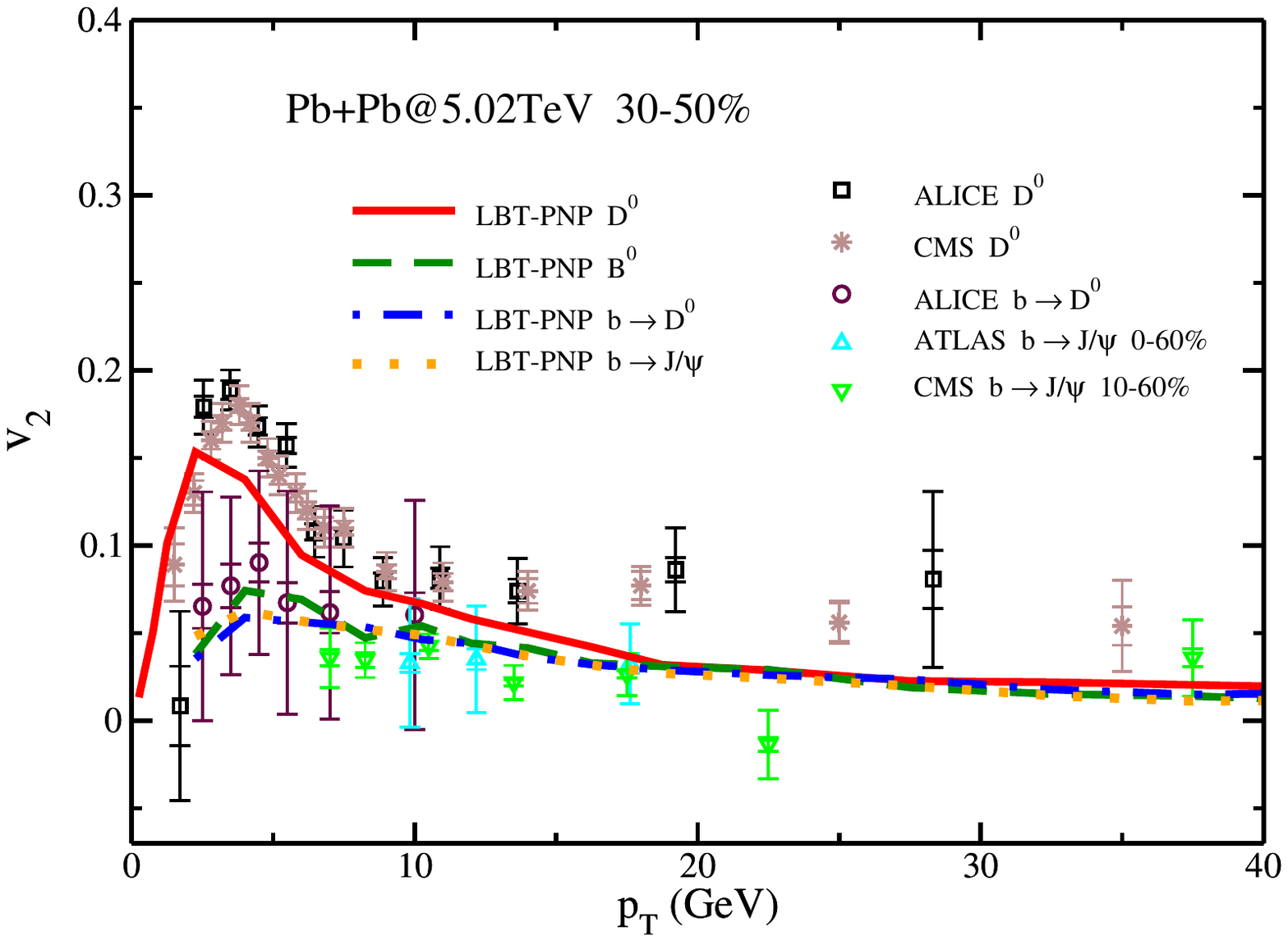}
\caption{(Color online) $R_\mathrm{AA}$ (0-10\% centrality) and $v_2$ (30-50\% centrality) of prompt $D^0$, $B$, non-prompt $D^0$ and non-prompt $J/\Psi$ as functions of $p_\mathrm{T}$ in Pb+Pb collisions at $\sqrt{s_\mathrm{NN}}=5.02$~TeV, compared to the available ALICE, ATLAS and CMS data~\cite{ALICE:2021rxa,CMS:2017qjw,ALICE:2022tji,ATLAS:2018hqe,ALICE:2023hou,ALICE:2020iug,CMS:2020bnz,ALICE:2023gjj,CMS:2022vfn,CMS:2023mtk,ATLAS:2018xms}.
} \label{fig6}
\end{figure*}

\begin{figure*}[tb]
\includegraphics[width=0.495\linewidth]{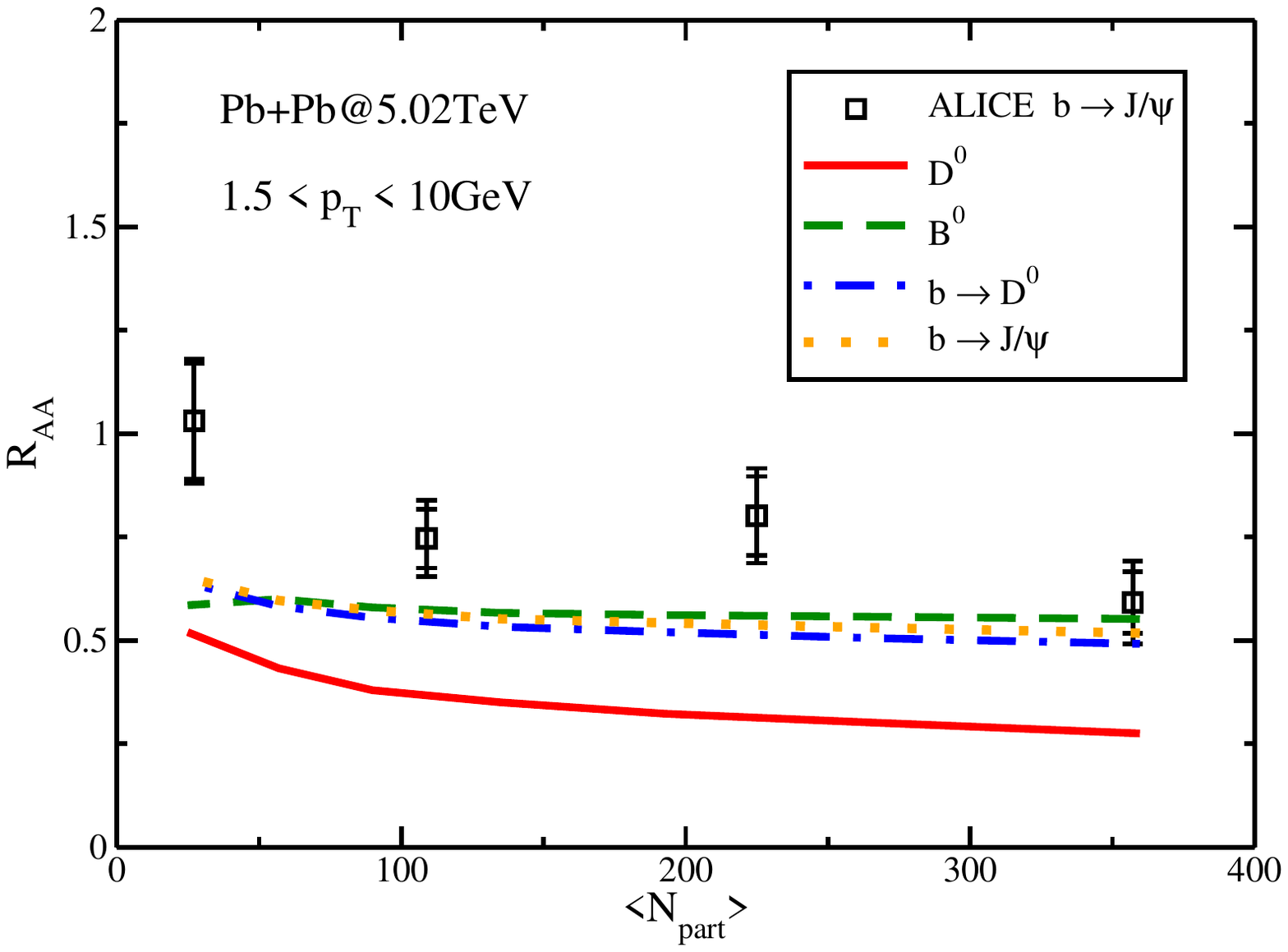}
\includegraphics[width=0.495\linewidth]{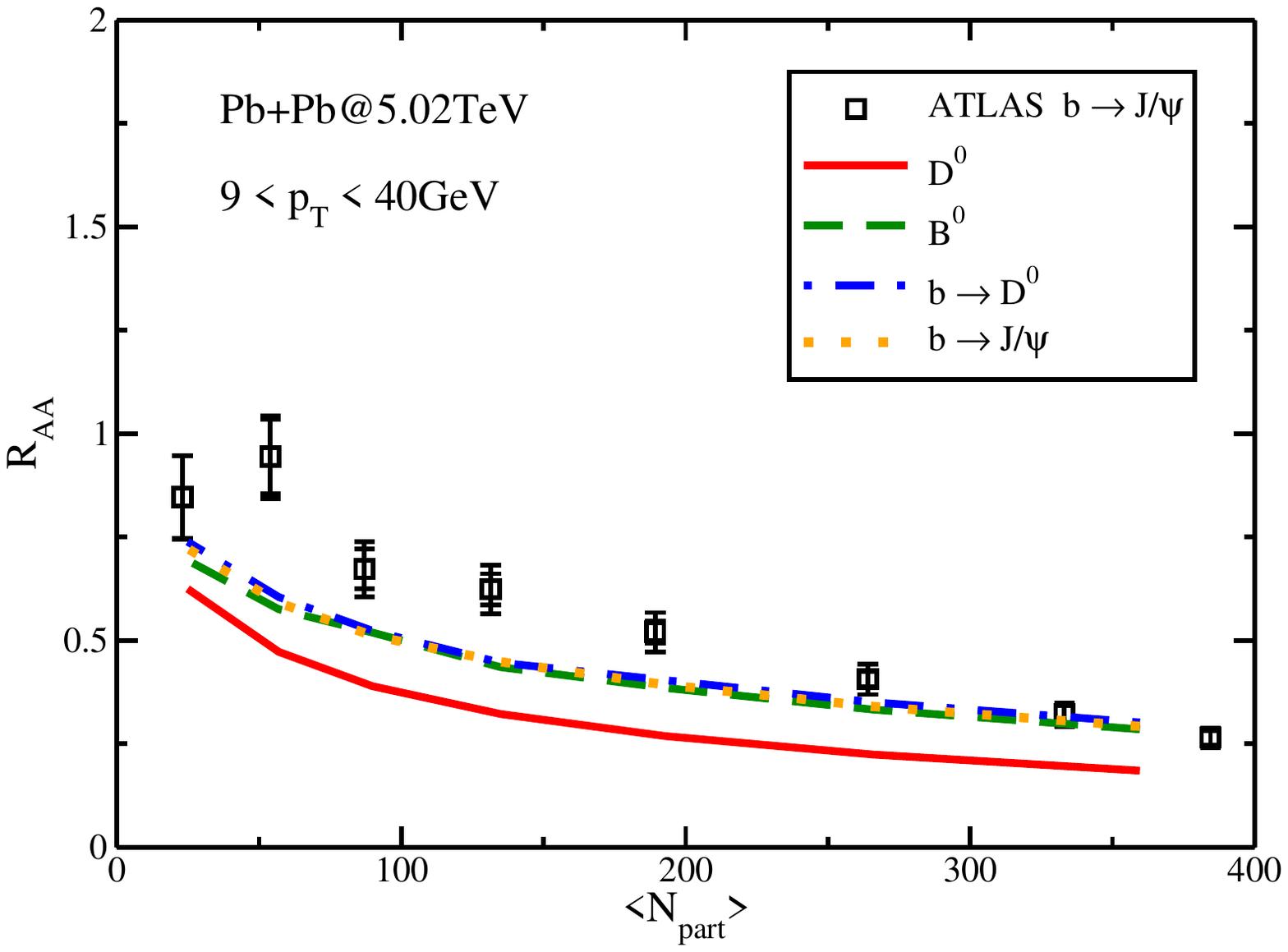}
\caption{(Color online) $R_\mathrm{AA}$ of prompt $D^0$, $B$, non-prompt $D^0$ and non-prompt $J/\Psi$, integrated over $1.5<p_\mathrm{T}<10$~GeV (left panel) and $9<p_\mathrm{T}<40$~GeV (right panel), as functions of the participant nucleon number in Pb+Pb collisions at $\sqrt{s_\mathrm{NN}}=5.02$~TeV, compared to the available ALICE and ATLAS data~\cite{ALICE:2023hou,ATLAS:2018hqe}.
} \label{fig7}
\end{figure*}

\begin{figure*}[tb]
\includegraphics[width=0.495\linewidth]{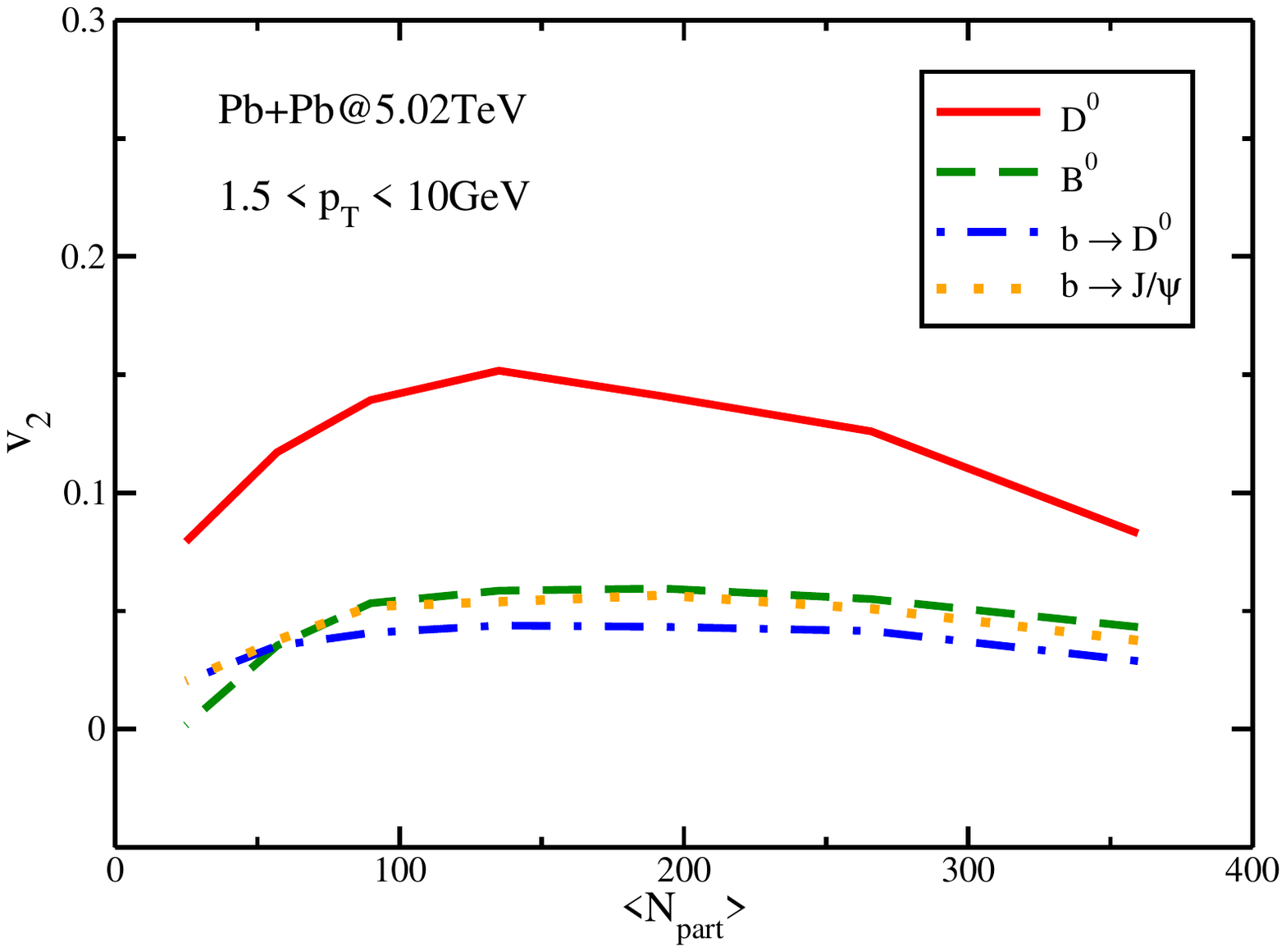}
\includegraphics[width=0.495\linewidth]{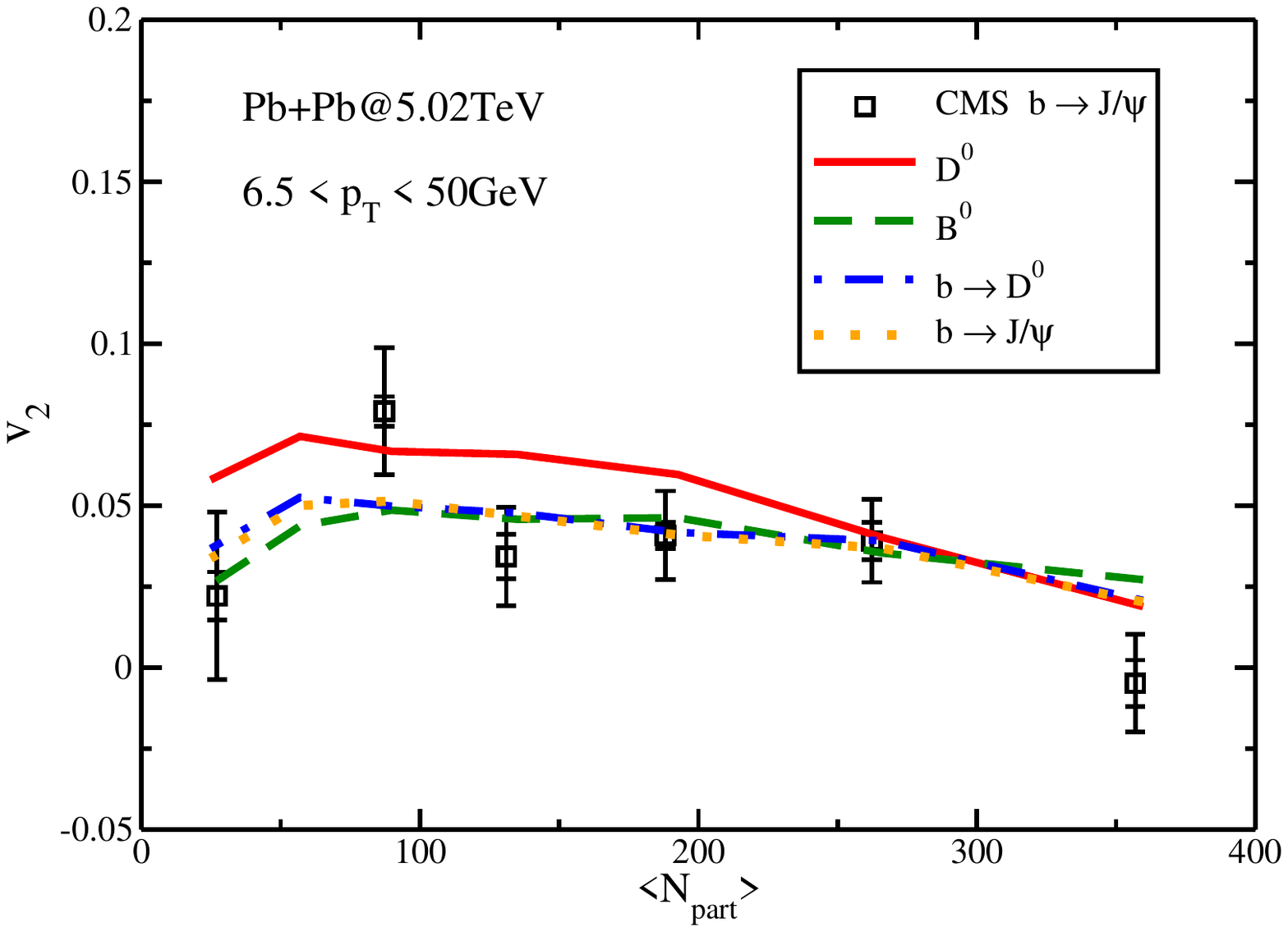}
\caption{(Color online) $v_2$ of prompt $D^0$, $B$, non-prompt $D^0$ and non-prompt $J/\Psi$, integrated over $1.5<p_\mathrm{T}<10$~GeV (left panel) and $6.5<p_\mathrm{T}<50$~GeV (right panel), as functions of the participant nucleon number in Pb+Pb collisions at $\sqrt{s_\mathrm{NN}}=5.02$~TeV, compared to the available CMS data~\cite{CMS:2023mtk}.
} \label{fig8}
\end{figure*}

\section{Numerical results}
\label{sec:results}

In this section, we present our numerical results on the nuclear modification factor ($R_\mathrm{AA}$) and elliptic flow coefficient ($v_2$) of non-prompt $D^0$ and $J/\psi$ in Pb+Pb collisions at $\sqrt{s_\mathrm{NN}} = 5.02$~TeV at the LHC.

Before showing results for non-prompt $D^0$ and $J/\psi$, we first show in Fig.~\ref{fig1} the $R_\mathrm{AA}$ and $v_2$ of prompt $D^0$ mesons as functions of transverse momentum $p_\mathrm{T}$ in Pb+Pb collisions at $\sqrt{s_\mathrm{NN}} = 5.02$~TeV for three different centrality classes: central (0-10\%), mid-central (30-50\%) and peripheral (60-80\%) collisions.
The ALICE and CMS data on 0-10\% and 30-50\% centralities are shown for comparison.
Our model can provide a reasonable description of the LHC data on the $D^0$ meson $R_\mathrm{AA}$ and $v_2$.
One can see that both $R_\mathrm{AA}$ and $v_2$ show strong centrality dependence.
From central to mid-central to peripheral collisions, the quenching of the prompt $D$ mesons decreases ($R_\mathrm{AA}$ becomes larger at large $p_\mathrm{T}$) due to the decreasing system size, whereas the prompt $D$ meson $v_2$ first increases due to the increasing eccentricity of QGP medium, and then decreases due to the decreasing energy loss through a smaller system.
Meanwhile, both $R_\mathrm{AA}$ and $v_2$ show strong $p_\mathrm{T}$ dependence.
The prompt $D$ meson $R_\mathrm{AA}$ first exhibits a bump structure at low $p_\mathrm{T}$ and then increases with $p_\mathrm{T}$ at high $p_\mathrm{T}$, whereas the prompt $D$ meson $v_2$ first increases and then decreases. These structures originate from the combined effect of heavy quark spectrum, energy loss and hadronization, together with the QGP flow. At large $p_\mathrm{T}$, the quenching of prompt $D$ mesons mainly comes from charm quark energy loss inside the QGP, and the $v_2$ of prompt $D$ mesons mainly results from the anisotropic energy loss along different directions through a geometrically asymmetric medium. The decreasing fractional energy loss and the flatter $p_\mathrm{T}$ spectrum of charm quarks at high $p_\mathrm{T}$ lead to the increasing $R_\mathrm{AA}$ and decreasing $v_2$ of prompt $D$ mesons with $p_\mathrm{T}$.
At low to intermediate $p_\mathrm{T}$, the strong non-perturbative interactions can quickly drive heavy quarks towards thermal equilibrium with the QGP, and therefore the motion of heavy quarks are strongly affected by the QGP flow. In addition, the coalescence between heavy quarks and the medium partons dominates the heavy flavor hadron formation at low to intermediate $p_\mathrm{T}$, which further enhances the QGP flow effect on the prompt $D$ meson spectrum. These lead to the flow bump of the prompt $D$ meson $R_\mathrm{AA}$, and the increasing $D$ meson $v_2$ at low $p_\mathrm{T}$, similar to the behaviors of light flavor hadron $R_\mathrm{AA}$ and $v_2$ observed at low $p_\mathrm{T}$.

Figure~\ref{fig2} shows our prediction for the $R_\mathrm{AA}$ and $v_2$ of $B$ mesons as functions of $p_\mathrm{T}$ in Pb+Pb collisions at $\sqrt{s_\mathrm{NN}}=5.02$~TeV for 0-10\%, 30-50\% and 60-80\% centrality classes. Here, results for $B$ mesons include contributions from $B^+$ and $B^-$ in our simulation. Similar to previous results for prompt $D$ mesons, the $B$ meson $R_\mathrm{AA}$ and $v_2$ also show strong centrality dependence.
Moving from central to mid-central to peripheral collisions, the quenching of $B$ mesons becomes weaker due to the decreasing system size, while the $B$ meson $v_2$ first increases and then decreases due to the combined effect of medium eccentricity and size.
For the same centrality class, the $B$ meson $R_\mathrm{AA}$ is larger than the prompt $D$ meson $R_\mathrm{AA}$ at low $p_\mathrm{T}$, while the former is comparable to the latter or even slightly smaller than the latter at high $p_\mathrm{T}$. This is due to the non-trivial mass hierarchy of quark energy loss within our LBT-PNP model, as discussed in an earlier study~\cite{Dang:2023tmb}: $b$-quarks lose less energy than $c$-quarks at low $p_\mathrm{T}$ due to the ``dead cone effect" in the gluon emission process, while the former may lose more energy than the latter at high $p_\mathrm{T}$ due to the string interaction implemented in our model. Within the same centrality class, the $v_2$ of $B$ mesons is smaller than that of prompt $D$ mesons at low $p_\mathrm{T}$. At high $p_\mathrm{T}$, they are both small.

To further study the quenching and flow of $B$ mesons, Recently, ALICE, ATLAS and CMS Collaborations have measured the $R_\mathrm{AA}$ and $v_2$ of non-prompt $D^0$ and $J/\psi$, decayed from bottom hadrons~\cite{CMS:2018bwt, ALICE:2022tji, ALICE:2023gjj, ATLAS:2018xms, ATLAS:2018hqe, CMS:2023mtk, ALICE:2023hou}.
In Fig.~\ref{fig3}, we show the $R_\mathrm{AA}$ and $v_2$ for non-prompt $D^0$ as functions of $p_\mathrm{T}$ in central (0-10\%), mid-central (30-50\%) and peripheral (60-80\%) Pb+Pb collisions at $\sqrt{s_\mathrm{NN}}=5.02$~TeV, compared to the ALICE and CMS data for 0-10\% and 30-50\% centralities.
One can see that our model provides a reasonable description of the non-prompt $D^0$ meson $R_\mathrm{AA}$ and $v_2$ at high $p_\mathrm{T}$.
At low $p_\mathrm{T}$, our model underestimates both $R_\mathrm{AA}$ and $v_2$, which may result from deficiencies in evaluating non-perturbative processes at low $p_\mathrm{T}$, including the initial heavy quark spectra, string interactions between heavy quarks and the QGP, and hadronization of heavy quarks.
Similar to prompt $D$ and $B$ mesons, the $R_\mathrm{AA}$ and $v_2$ of non-prompt $D^0$ mesons here also show strong dependence on centrality.
In the same centrality class, the $R_\mathrm{AA}$ and $v_2$ of non-prompt $D^0$ show similar behaviors as those of $B$ mesons, except for some $p_\mathrm{T}$ shift during the decay from $B$ mesons to $D$ mesons.

In Fig.~\ref{fig4}, we show the $R_\mathrm{AA}$ of non-prompt $J/\psi$ as a function of $p_\mathrm{T}$ in 0-80\% Pb+Pb collisions at $\sqrt{s_\mathrm{NN}}=5.02$~TeV and the $v_2$ of non-prompt $J/\psi$ in 10-60\% collisions.
Our model provides a good description of the corresponding ATLAS data on $R_\mathrm{AA}$ and the CMS data on $v_2$ of non-prompt $J/\psi$.
Both the $R_\mathrm{AA}$ and $v_2$ results here are obtained from averaging over a large interval of centrality. They do not show a strong dependence on $p_\mathrm{T}$.

In order to investigate the centrality dependence of quenching and flow of non-prompt $J/\psi$, we present in Fig.~\ref{fig5} the $R_\mathrm{AA}$ and $v_2$ of non-prompt $J/\psi$ as functions of $p_\mathrm{T}$ in central (0-10\%), mid-central (30-50\%) and peripheral (60-80\%) Pb+Pb collisions at $\sqrt{s_\mathrm{NN}}=5.02$~TeV, compared to the available data from the ALICE, ATLAS and CMS collaborations.
One can see that our model provides a reasonable description of the corresponding data within comparable centrality bins.
Similar to previous results of prompt $D$ mesons, $B$ mesons and non-prompt $D^0$, the $R_\mathrm{AA}$ and $v_2$ of non-prompt $J/\psi$ show strong centrality dependence.

In Fig.~\ref{fig6}, we directly compare the quenching and flow between different species of heavy flavor particles presented above, left panel for the $R_\mathrm{AA}$ of prompt $D$, $B$, non-prompt $D^0$ and non-prompt $J/\psi$ in central (0-10\%) Pb+Pb collisions at $\sqrt{s_\mathrm{NN}}=5.02$~TeV, and right panel for their $v_2$ in mid-central (30-50\%) collisions, in comparison to the available ALICE, ATLAS and CMS data. As discussed earlier, the $R_\mathrm{AA}$ of direct $D$ mesons is smaller than that of $B$ mesons at low $p_\mathrm{T}$, while the inverse order is seen at high $p_\mathrm{T}$. This is because of the opposite mass dependences of quark energy loss at low $p_\mathrm{T}$ and high $p_\mathrm{T}$ within our LBT-PNP model. In mid-central collisions, the $v_2$ of direct $D$ mesons appear larger than that of $B$ mesons. The $R_\mathrm{AA}$ of non-prompt $D^0$ and non-prompt $J/\psi$ are smaller than that of $B$ mesons at low $p_\mathrm{T}$, but larger at high $p_\mathrm{T}$, due to the $p_\mathrm{T}$ shift during the decay of $B$ mesons. The $v_2$ of non-prompt $D^0$ and non-prompt $J/\psi$ are smaller than that of $B$ mesons at low $p_\mathrm{T}$, but comparable at high $p_\mathrm{T}$. No apparent difference in $R_\mathrm{AA}$ and $v_2$ is observed between non-prompt $D^0$ and non-prompt $J/\psi$ due to their similar decay functions from $B$ mesons.

In Fig.~\ref{fig7}, we compare the $p_\mathrm{T}$-integrated $R_\mathrm{AA}$ between prompt $D^0$, $B$, non-prompt $D^0$ and non-prompt $J/\psi$ in Pb+Pb collisions at $\sqrt{s_\mathrm{NN}}=5.02$~TeV, left panel for $1.5<p_\mathrm{T}<10$~GeV and right panel for $9<p_\mathrm{T}<40$~GeV. At higher $p_\mathrm{T}$ region (right panel), a clear participant number ($N_\mathrm{part}$), or centrality, dependence of $R_\mathrm{AA}$ can be seen for these heavy flavor particles: stronger energy loss of heavy quarks in more central collisions (larger $N_\mathrm{part}$) leads to smaller $R_\mathrm{AA}$. This trend is not apparent at low $p_\mathrm{T}$ (left panel), where in addition to parton energy loss, the hadronization process and the QGP flow also significantly affect the final state hadron spectra. Within the $p_\mathrm{T}$ range we explore here, the $R_\mathrm{AA}$ of direct $D$ mesons is smaller than that of $B$ mesons. No obvious difference is observed between $B$ mesons, non-prompt $D^0$ and non-prompt $J/\psi$ in these $p_\mathrm{T}$-integrated $R_\mathrm{AA}$. Since our current model underestimates the heavy flavor $R_\mathrm{AA}$ at low $p_\mathrm{T}$, as shown in previous $p_\mathrm{T}$-dependent $R_\mathrm{AA}$ results, our result on the $p_\mathrm{T}$-integrated $R_\mathrm{AA}$ of non-prompt $J/\psi$ here is also lower than the available data from the ALICE and ATLAS collaborations. The agreement becomes better as the $p_\mathrm{T}$ range becomes higher.

In the end, we compare in Fig.~\ref{fig8} the $p_\mathrm{T}$-integrated $v_2$ between prompt $D^0$, $B$, non-prompt $D^0$ and non-prompt $J/\psi$ in Pb+Pb collisions at $\sqrt{s_\mathrm{NN}}=5.02$~TeV, left panel for $1.5<p_\mathrm{T}<10$~GeV and right panel for $6.5<p_\mathrm{T}<50$~GeV. Moving from central to mid-central to peripheral collisions, or as $N_\mathrm{part}$ becomes smaller, the elliptic flow coefficients first increase and then decrease due to the combined effect of medium eccentricity and medium size. At low $p_\mathrm{T}$, the prompt $D^0$ mesons have much larger $v_2$ than $B$ mesons, non-prompt $D^0$ and non-prompt $J/\psi$. This difference becomes smaller at higher $p_\mathrm{T}$. Compared to the available data from the CMS collaboration, our model provides a reasonable description of the $v_2$ of non-prompt $J/\psi$.


\section{Summary}
\label{sec:summary}

Within the linear Boltzmann transport model that includes both string and Yukawa types of interactions between heavy quarks and the QGP, we study the dynamics of bottom quarks in Pb+Pb collisions at $\sqrt{s_\mathrm{NN}}=5.02$~TeV at the LHC via the nuclear modification factors and elliptic flow coefficients of $B$ mesons, bottom decayed $D^0$ and $J/\psi$. Compared to direct $D$ mesons, $B$ mesons show larger $R_\mathrm{AA}$ at low $p_\mathrm{T}$ but slightly smaller $R_\mathrm{AA}$ at high $p_\mathrm{T}$, indicating weaker energy loss of heavier quarks at low $p_\mathrm{T}$ but a possible inverse order at high $p_\mathrm{T}$ within our LBT-PNP model. Both $R_\mathrm{AA}$ and $v_2$ of $D$ and $B$ mesons show strong $p_\mathrm{T}$ and centrality dependences. At high $p_\mathrm{T}$, the heavy meson $R_\mathrm{AA}$ increases with $p_\mathrm{T}$ due to the decreasing fractional energy loss and the flatter $p_\mathrm{T}$ spectra of heavy quarks at higher $p_\mathrm{T}$. A bump structure of $R_\mathrm{AA}$ can be observed at low $p_\mathrm{T}$, resulting from the QGP flow effect and the coalescence process of heavy quark hadronization. The heavy meson $v_2$ is driven by the anisotropic QGP flow at low $p_\mathrm{T}$, while driven by anisotropic energy loss through different directions at high $p_\mathrm{T}$, and therefore, first increases and then decreases as $p_\mathrm{T}$ becomes larger. From central to mid-central to peripheral collisions, the heavy meson $R_\mathrm{AA}$ becomes larger due to weaker energy loss through a smaller QGP system. On the other hand, their $v_2$ first increases due to larger medium eccentricity, and then decreases due to smaller medium size. Non-prompt $D^0$ and non-prompt $J/\psi$ show very similar $R_\mathrm{AA}$ and $v_2$ to $B$ mesons, except for a shift towards the lower $p_\mathrm{T}$ region. Compared to the available ALICE, ATLAS and CMS data, our model provides a reasonable description of the nuclear modification factors and elliptic flow coefficients of prompt $D$ mesons, non-prompt $D^0$ and non-prompt $J/\psi$, except for some deviation at low $p_\mathrm{T}$ due to possible inaccurate description of initial spectrum of heavy quarks, their interaction with the QGP and their hadronization process in this non-perturbative region. Therefore, studying non-prompt $D^0$ and $J/\psi$ provides a supplementary way for better understanding heavy quark dynamics in relativistic heavy-ion collisions.


\section*{Acknowledgments}

This work is supported in part by the National Natural Science Foundation of China (NSFC) under Grant Nos. 12225503, 11890710, 11890711, 11935007, 12175122 and 2021-867.  W.-J.~X. is supported in part by China Postdoctoral Science Foundation under Grant No. 2023M742099. Some of the calculations were performed in the Nuclear Science Computing Center at Central China Normal University (NSC$^3$), Wuhan, Hubei, China.

\bibliographystyle{plain}
\bibliographystyle{h-physrev5}
\bibliography{refs_GYQ}

\begin{thebibliography}{100}

\bibitem{Gyulassy:2004zy}
M.~Gyulassy and L.~McLerran,
\newblock Nucl. Phys. A {\bf 750}, 30 (2005), arXiv:nucl-th/0405013.

\bibitem{Muller:2012zq}
B.~Muller, J.~Schukraft, and B.~Wyslouch,
\newblock Ann. Rev. Nucl. Part. Sci. {\bf 62}, 361 (2012), arXiv:1202.3233.

\bibitem{Romatschke:2017ejr}
P.~Romatschke and U.~Romatschke,
\newblock {\em {Relativistic Fluid Dynamics In and Out of Equilibrium,
  }}Cambridge Monographs on Mathematical Physics (Cambridge University Press,
  2019), arXiv:1712.05815.

\bibitem{Rischke:1995ir}
D.~H. Rischke, S.~Bernard, and J.~A. Maruhn,
\newblock Nucl. Phys. A {\bf 595}, 346 (1995), arXiv:nucl-th/9504018.

\bibitem{Heinz:2013th}
U.~Heinz and R.~Snellings,
\newblock Ann. Rev. Nucl. Part. Sci. {\bf 63}, 123 (2013), arXiv:1301.2826.

\bibitem{Gale:2013da}
C.~Gale, S.~Jeon, and B.~Schenke,
\newblock Int. J. Mod. Phys. A {\bf 28}, 1340011 (2013), arXiv:1301.5893.

\bibitem{Huovinen:2013wma}
P.~Huovinen,
\newblock Int. J. Mod. Phys. E {\bf 22}, 1330029 (2013), arXiv:1311.1849.

\bibitem{Wang:1991xy}
X.-N. Wang and M.~Gyulassy,
\newblock Phys. Rev. Lett. {\bf 68}, 1480 (1992).

\bibitem{Gyulassy:2003mc}
M.~Gyulassy, I.~Vitev, X.-N. Wang, and B.-W. Zhang,
\newblock (2003), arXiv:nucl-th/0302077.

\bibitem{Majumder:2010qh}
A.~Majumder and M.~Van~Leeuwen,
\newblock Prog. Part. Nucl. Phys. {\bf 66}, 41 (2011), arXiv:1002.2206.

\bibitem{Qin:2015srf}
G.-Y. Qin and X.-N. Wang,
\newblock Int. J. Mod. Phys. {\bf E24}, 1530014 (2015), arXiv:1511.00790.

\bibitem{Blaizot:2015lma}
J.-P. Blaizot and Y.~Mehtar-Tani,
\newblock Int. J. Mod. Phys. E {\bf 24}, 1530012 (2015), arXiv:1503.05958.

\bibitem{Cao:2020wlm}
S.~Cao and X.-N. Wang,
\newblock Rept. Prog. Phys. {\bf 84}, 024301 (2021), arXiv:2002.04028.

\bibitem{Cao:2022odi}
S.~Cao and G.-Y. Qin,
\newblock Ann. Rev. Nucl. Part. Sci. {\bf 73}, 205 (2023), arXiv:2211.16821.

\bibitem{Dong:2019byy}
X.~Dong, Y.-J. Lee, and R.~Rapp,
\newblock Ann. Rev. Nucl. Part. Sci. {\bf 69}, 417 (2019), arXiv:1903.07709.

\bibitem{Andronic:2015wma}
A.~Andronic {\em et~al.},
\newblock Eur. Phys. J. {\bf C76}, 107 (2016), arXiv:1506.03981.

\bibitem{He:2022ywp}
M.~He, H.~van Hees, and R.~Rapp,
\newblock Prog. Part. Nucl. Phys. {\bf 130}, 104020 (2023), arXiv:2204.09299.

\bibitem{Cao:2013ita}
S.~Cao, G.-Y. Qin, and S.~A. Bass,
\newblock Phys. Rev. {\bf C88}, 044907 (2013), arXiv:1308.0617.

\bibitem{Cao:2016gvr}
S.~Cao, T.~Luo, G.-Y. Qin, and X.-N. Wang,
\newblock Phys. Rev. C {\bf 94}, 014909 (2016), arXiv:1605.06447.

\bibitem{Xing:2019xae}
W.-J. Xing, S.~Cao, G.-Y. Qin, and H.~Xing,
\newblock Phys. Lett. B {\bf 805}, 135424 (2020), arXiv:1906.00413.

\bibitem{Liu:2021dpm}
F.-L. Liu {\em et~al.},
\newblock (2021), arXiv:2107.11713.

\bibitem{STAR:2014wif}
STAR, L.~Adamczyk {\em et~al.},
\newblock Phys. Rev. Lett. {\bf 113}, 142301 (2014), arXiv:1404.6185,
\newblock [Erratum: Phys.Rev.Lett. 121, 229901 (2018)].

\bibitem{CMS:2017qjw}
CMS, A.~M. Sirunyan {\em et~al.},
\newblock Phys. Lett. B {\bf 782}, 474 (2018), arXiv:1708.04962.

\bibitem{STAR:2018zdy}
STAR, J.~Adam {\em et~al.},
\newblock Phys. Rev. C {\bf 99}, 034908 (2019), arXiv:1812.10224.

\bibitem{Gossiaux:2006yu}
P.~B. Gossiaux, V.~Guiho, and J.~Aichelin,
\newblock J. Phys. G {\bf 32}, S359 (2006).

\bibitem{Qin:2009gw}
G.-Y. Qin and A.~Majumder,
\newblock Phys. Rev. Lett. {\bf 105}, 262301 (2010), arXiv:0910.3016.

\bibitem{Uphoff:2011ad}
J.~Uphoff, O.~Fochler, Z.~Xu, and C.~Greiner,
\newblock Phys. Rev. {\bf C84}, 024908 (2011), arXiv:1104.2295.

\bibitem{Young:2011ug}
C.~Young, B.~Schenke, S.~Jeon, and C.~Gale,
\newblock Phys. Rev. {\bf C86}, 034905 (2012), arXiv:1111.0647.

\bibitem{Alberico:2011zy}
W.~Alberico {\em et~al.},
\newblock Eur. Phys. J. {\bf C71}, 1666 (2011), arXiv:1101.6008.

\bibitem{Fochler:2013epa}
O.~Fochler, J.~Uphoff, Z.~Xu, and C.~Greiner,
\newblock Phys. Rev. {\bf D88}, 014018 (2013), arXiv:1302.5250.

\bibitem{Nahrgang:2013saa}
M.~Nahrgang, J.~Aichelin, P.~B. Gossiaux, and K.~Werner,
\newblock Phys. Rev. {\bf C90}, 024907 (2014), arXiv:1305.3823.

\bibitem{Djordjevic:2013xoa}
M.~Djordjevic and M.~Djordjevic,
\newblock Phys. Lett. B {\bf 734}, 286 (2014), arXiv:1307.4098.

\bibitem{Cao:2015hia}
S.~Cao, G.-Y. Qin, and S.~A. Bass,
\newblock Phys. Rev. {\bf C92}, 024907 (2015), arXiv:1505.01413.

\bibitem{Das:2015ana}
S.~K. Das, F.~Scardina, S.~Plumari, and V.~Greco,
\newblock Phys. Lett. B {\bf 747}, 260 (2015), arXiv:1502.03757.

\bibitem{Song:2015ykw}
T.~Song, H.~Berrehrah, D.~Cabrera, W.~Cassing, and E.~Bratkovskaya,
\newblock Phys. Rev. C {\bf 93}, 034906 (2016), arXiv:1512.00891.

\bibitem{Kang:2016ofv}
Z.-B. Kang, F.~Ringer, and I.~Vitev,
\newblock JHEP {\bf 03}, 146 (2017), arXiv:1610.02043.

\bibitem{Prado:2016szr}
C.~A.~G. Prado {\em et~al.},
\newblock Phys. Rev. {\bf C96}, 064903 (2017), arXiv:1611.02965.

\bibitem{Xu:2017obm}
Y.~Xu, J.~E. Bernhard, S.~A. Bass, M.~Nahrgang, and S.~Cao,
\newblock Phys. Rev. C {\bf 97}, 014907 (2018), arXiv:1710.00807.

\bibitem{Liu:2017qah}
S.~Y.~F. Liu and R.~Rapp,
\newblock Phys. Rev. {\bf C97}, 034918 (2018), arXiv:1711.03282.

\bibitem{Rapp:2018qla}
A.~Beraudo {\em et~al.},
\newblock Nucl. Phys. {\bf A979}, 21 (2018), arXiv:1803.03824.

\bibitem{Cao:2018ews}
S.~Cao {\em et~al.},
\newblock Phys. Rev. {\bf C99}, 054907 (2019), arXiv:1809.07894.

\bibitem{Li:2018izm}
S.~Li, C.~Wang, X.~Yuan, and S.~Feng,
\newblock Phys. Rev. {\bf C98}, 014909 (2018), arXiv:1803.01508.

\bibitem{Ke:2018tsh}
W.~Ke, Y.~Xu, and S.~A. Bass,
\newblock Phys. Rev. C {\bf 98}, 064901 (2018), arXiv:1806.08848.

\bibitem{Li:2019wri}
S.~Li, C.~Wang, R.~Wan, and J.~Liao,
\newblock Phys. Rev. C {\bf 99}, 054909 (2019), arXiv:1901.04600.

\bibitem{Katz:2019fkc}
R.~Katz, C.~A. Prado, J.~Noronha-Hostler, J.~Noronha, and A.~A. Suaide,
\newblock (2019), arXiv:1906.10768.

\bibitem{Li:2020kax}
S.-Q. Li, W.-J. Xing, F.-L. Liu, S.~Cao, and G.-Y. Qin,
\newblock Chin. Phys. C {\bf 44}, 114101 (2020), arXiv:2005.03330.

\bibitem{Chen:2021uar}
B.~Chen, L.~Wen, and Y.~Liu,
\newblock Phys. Lett. B {\bf 834}, 137448 (2022), arXiv:2111.08490.

\bibitem{Yang:2023rgb}
M.~Yang {\em et~al.},
\newblock Phys. Rev. C {\bf 107}, 054917 (2023), arXiv:2302.06179.

\bibitem{Liu:2023rfi}
F.-L. Liu, X.-Y. Wu, S.~Cao, G.-Y. Qin, and X.-N. Wang,
\newblock (2023), arXiv:2304.08787.

\bibitem{Ollitrault:1992bk}
J.-Y. Ollitrault,
\newblock Phys. Rev. D {\bf 46}, 229 (1992).

\bibitem{STAR:2000ekf}
STAR, K.~H. Ackermann {\em et~al.},
\newblock Phys. Rev. Lett. {\bf 86}, 402 (2001), arXiv:nucl-ex/0009011.

\bibitem{STAR:2001ksn}
STAR, C.~Adler {\em et~al.},
\newblock Phys. Rev. Lett. {\bf 87}, 182301 (2001), arXiv:nucl-ex/0107003.

\bibitem{ALICE:2010suc}
ALICE, K.~Aamodt {\em et~al.},
\newblock Phys. Rev. Lett. {\bf 105}, 252302 (2010), arXiv:1011.3914.

\bibitem{ALICE:2011ab}
ALICE, K.~Aamodt {\em et~al.},
\newblock Phys. Rev. Lett. {\bf 107}, 032301 (2011), arXiv:1105.3865.

\bibitem{Alver:2010gr}
B.~Alver and G.~Roland,
\newblock Phys. Rev. {\bf C81}, 054905 (2010), arXiv:1003.0194.

\bibitem{Qin:2010pf}
G.-Y. Qin, H.~Petersen, S.~A. Bass, and B.~Muller,
\newblock Phys. Rev. {\bf C82}, 064903 (2010), arXiv:1009.1847.

\bibitem{Moore:2004tg}
G.~D. Moore and D.~Teaney,
\newblock Phys. Rev. {\bf C71}, 064904 (2005), hep-ph/0412346.

\bibitem{He:2011qa}
M.~He, R.~J. Fries, and R.~Rapp,
\newblock Phys. Rev. C {\bf 86}, 014903 (2012), arXiv:1106.6006.

\bibitem{Xing:2021xwc}
W.-J. Xing, G.-Y. Qin, and S.~Cao,
\newblock Phys. Lett. B {\bf 838}, 137733 (2023), arXiv:2112.15062.

\bibitem{Plumari:2017ntm}
S.~Plumari, V.~Minissale, S.~K. Das, G.~Coci, and V.~Greco,
\newblock Eur. Phys. J. {\bf C78}, 348 (2018), arXiv:1712.00730.

\bibitem{He:2019vgs}
M.~He and R.~Rapp,
\newblock Phys. Rev. Lett. {\bf 124}, 042301 (2020), arXiv:1905.09216.

\bibitem{Cho:2019lxb}
S.~Cho, K.-J. Sun, C.~M. Ko, S.~H. Lee, and Y.~Oh,
\newblock Phys. Rev. C {\bf 101}, 024909 (2020), arXiv:1905.09774.

\bibitem{Cao:2019iqs}
S.~Cao {\em et~al.},
\newblock Phys. Lett. B {\bf 807}, 135561 (2020), arXiv:1911.00456.

\bibitem{Zhao:2023nrz}
J.~Zhao {\em et~al.},
\newblock (2023), arXiv:2311.10621.

\bibitem{STAR:2017kkh}
STAR, L.~Adamczyk {\em et~al.},
\newblock Phys. Rev. Lett. {\bf 118}, 212301 (2017), arXiv:1701.06060.

\bibitem{CMS:2017vhp}
CMS, A.~M. Sirunyan {\em et~al.},
\newblock Phys. Rev. Lett. {\bf 120}, 202301 (2018), arXiv:1708.03497.

\bibitem{ALICE:2017pbx}
ALICE, S.~Acharya {\em et~al.},
\newblock Phys. Rev. Lett. {\bf 120}, 102301 (2018), arXiv:1707.01005.

\bibitem{ALICE:2020iug}
ALICE, S.~Acharya {\em et~al.},
\newblock Phys. Lett. B {\bf 813}, 136054 (2021), arXiv:2005.11131.

\bibitem{Kusina:2017gkz}
A.~Kusina, J.-P. Lansberg, I.~Schienbein, and H.-S. Shao,
\newblock Phys. Rev. Lett. {\bf 121}, 052004 (2018), arXiv:1712.07024.

\bibitem{Dokshitzer:2001zm}
Y.~L. Dokshitzer and D.~E. Kharzeev,
\newblock Phys. Lett. B {\bf 519}, 199 (2001), arXiv:hep-ph/0106202.

\bibitem{Armesto:2003jh}
N.~Armesto, C.~A. Salgado, and U.~A. Wiedemann,
\newblock Phys. Rev. D {\bf 69}, 114003 (2004), arXiv:hep-ph/0312106.

\bibitem{Zhang:2003wk}
B.-W. Zhang, E.~Wang, and X.-N. Wang,
\newblock Phys. Rev. Lett. {\bf 93}, 072301 (2004), arXiv:nucl-th/0309040.

\bibitem{Djordjevic:2003zk}
M.~Djordjevic and M.~Gyulassy,
\newblock Nucl. Phys. {\bf A733}, 265 (2004), arXiv:nucl-th/0310076.

\bibitem{Zhang:2018nie}
L.~Zhang, D.-F. Hou, and G.-Y. Qin,
\newblock Phys. Rev. {\bf C100}, 034907 (2019), arXiv:1812.11048.

\bibitem{CMS:2018bwt}
CMS, A.~M. Sirunyan {\em et~al.},
\newblock Phys. Rev. Lett. {\bf 123}, 022001 (2019), arXiv:1810.11102.

\bibitem{ALICE:2022tji}
ALICE, S.~Acharya {\em et~al.},
\newblock JHEP {\bf 12}, 126 (2022), arXiv:2202.00815.

\bibitem{ALICE:2023gjj}
ALICE, S.~Acharya {\em et~al.},
\newblock Eur. Phys. J. C {\bf 83}, 1123 (2023), arXiv:2307.14084.

\bibitem{ATLAS:2018xms}
ATLAS, M.~Aaboud {\em et~al.},
\newblock Eur. Phys. J. C {\bf 78}, 784 (2018), arXiv:1807.05198.

\bibitem{ATLAS:2018hqe}
ATLAS, M.~Aaboud {\em et~al.},
\newblock Eur. Phys. J. C {\bf 78}, 762 (2018), arXiv:1805.04077.

\bibitem{ATLAS:2020yxw}
ATLAS, G.~Aad {\em et~al.},
\newblock Phys. Lett. B {\bf 807}, 135595 (2020), arXiv:2003.03565.

\bibitem{ATLAS:2021xtw}
ATLAS, G.~Aad {\em et~al.},
\newblock Phys. Lett. B {\bf 829}, 137077 (2022), arXiv:2109.00411.

\bibitem{ALICE:2019nuy}
ALICE, S.~Acharya {\em et~al.},
\newblock Phys. Lett. B {\bf 804}, 135377 (2020), arXiv:1910.09110.

\bibitem{CMS:2023mtk}
CMS, A.~Tumasyan {\em et~al.},
\newblock JHEP {\bf 10}, 115 (2023), arXiv:2305.16928.

\bibitem{ALICE:2023hou}
ALICE, S.~Acharya {\em et~al.},
\newblock JHEP {\bf 02}, 066 (2024), arXiv:2308.16125.

\bibitem{He:2015pra}
Y.~He, T.~Luo, X.-N. Wang, and Y.~Zhu,
\newblock Phys. Rev. {\bf C91}, 054908 (2015), arXiv:1503.03313.

\bibitem{Cao:2017hhk}
S.~Cao, T.~Luo, G.-Y. Qin, and X.-N. Wang,
\newblock Phys. Lett. {\bf B777}, 255 (2018), arXiv:1703.00822.

\bibitem{Combridge:1978kx}
B.~Combridge,
\newblock Nucl. Phys. {\bf B151}, 429 (1979).

\bibitem{Svetitsky:1987gq}
B.~Svetitsky,
\newblock Phys. Rev. {\bf D37}, 2484 (1988).

\bibitem{GolamMustafa:1997id}
M.~Mustafa, D.~Pal, and D.~Kumar~Srivastava,
\newblock Phys. Rev. {\bf C57}, 889 (1998), arXiv:nucl-th/9706001.

\bibitem{Liu:2016ysz}
S.~Y.~F. Liu and R.~Rapp,
\newblock Eur. Phys. J. A {\bf 56}, 44 (2020), arXiv:1612.09138.

\bibitem{Guo:2000nz}
X.-F. Guo and X.-N. Wang,
\newblock Phys. Rev. Lett. {\bf 85}, 3591 (2000), arXiv:hep-ph/0005044.

\bibitem{Majumder:2009ge}
A.~Majumder,
\newblock Phys. Rev. {\bf D85}, 014023 (2012), arXiv:0912.2987.

\bibitem{Pang:2012he}
L.~Pang, Q.~Wang, and X.-N. Wang,
\newblock Phys. Rev. {\bf C86}, 024911 (2012), arXiv:1205.5019.

\bibitem{Pang:2018zzo}
L.-G. Pang, H.~Petersen, and X.-N. Wang,
\newblock Phys. Rev. C {\bf 97}, 064918 (2018), arXiv:1802.04449.

\bibitem{Wu:2018cpc}
X.-Y. Wu, L.-G. Pang, G.-Y. Qin, and X.-N. Wang,
\newblock Phys. Rev. C {\bf 98}, 024913 (2018), arXiv:1805.03762.

\bibitem{Wu:2021fjf}
X.-Y. Wu, G.-Y. Qin, L.-G. Pang, and X.-N. Wang,
\newblock Phys. Rev. C {\bf 105}, 034909 (2022), arXiv:2107.04949.

\bibitem{Cacciari:2001td}
M.~Cacciari, S.~Frixione, and P.~Nason,
\newblock JHEP {\bf 03}, 006 (2001), arXiv:hep-ph/0102134.

\bibitem{Cacciari:2012ny}
M.~Cacciari {\em et~al.},
\newblock JHEP {\bf 10}, 137 (2012), arXiv:1205.6344.

\bibitem{Cacciari:2015fta}
M.~Cacciari, M.~L. Mangano, and P.~Nason,
\newblock Eur. Phys. J. {\bf C75}, 610 (2015), arXiv:1507.06197.

\bibitem{Dulat:2015mca}
S.~Dulat {\em et~al.},
\newblock Phys. Rev. {\bf D93}, 033006 (2016), arXiv:1506.07443.

\bibitem{Eskola:2016oht}
K.~J. Eskola, P.~Paakkinen, H.~Paukkunen, and C.~A. Salgado,
\newblock Eur. Phys. J. {\bf C77}, 163 (2017), arXiv:1612.05741.

\bibitem{Sjostrand:2006za}
T.~Sjostrand, S.~Mrenna, and P.~Z. Skands,
\newblock JHEP {\bf 0605}, 026 (2006), arXiv:hep-ph/0603175.

\bibitem{ALICE:2021rxa}
ALICE, S.~Acharya {\em et~al.},
\newblock JHEP {\bf 01}, 174 (2022), arXiv:2110.09420.

\bibitem{CMS:2020bnz}
CMS, A.~M. Sirunyan {\em et~al.},
\newblock Phys. Lett. B {\bf 816}, 136253 (2021), arXiv:2009.12628.

\bibitem{CMS:2022vfn}
CMS, A.~Tumasyan {\em et~al.},
\newblock Phys. Lett. B {\bf 850}, 138389 (2024), arXiv:2212.01636.

\bibitem{Dang:2023tmb}
Y.~Dang, W.-J. Xing, S.~Cao, and G.-Y. Qin,
\newblock (2023), arXiv:2307.14808.

\end{thebibliography}
\end{document}